\documentclass[11pt,reqno]{amsart}
\usepackage{paralist}
\usepackage{bbm}
\usepackage[lite]{amsrefs}
\usepackage{amssymb}
\usepackage[all,cmtip]{xy}
\usepackage{amsmath}
\usepackage{amsthm}
\usepackage{amsfonts}
\usepackage{hyperref}
\usepackage{enumerate}
\usepackage[titletoc]{appendix}
\usepackage{fancyhdr}
\usepackage[dvipsnames]{xcolor}
\usepackage{comment}
\usepackage{tikz-cd}
\usepackage{tikzpagenodes}
\usepackage{caption}
\usepackage{float}
\usetikzlibrary[patterns]
\usepackage{setspace}

\allowdisplaybreaks

\newcommand{\g}{\mathfrak{g}}
\newcommand{\n}{\mathfrak{n}}

\newcommand{\Z}{\mathbb{Z}}

\newcommand{\C}{\mathbb{C}}

\newcommand{\Ker}{\operatorname{Ker}}
\newcommand{\Img}{\operatorname{Im}}

\newcommand{\ad}{\operatorname{ad}}

\newcommand{\NO}[1]{:\!#1\!:}
\newcommand{\vac}{|0\rangle}

\newcommand{\e}{\mathrm{e}}
\newcommand{\h}{\mathfrak{h}}

\newcommand{\DistTo}{\xrightarrow{
   \,\smash{\raisebox{-0.45ex}{\ensuremath{\scriptstyle\sim}}}\,}}

 \newtheorem{thm}{Theorem}[section]

 \newtheorem{cor}[thm]{Corollary}
 
 \newtheorem{prop}[thm]{Proposition}
 
 \newtheorem{defn}[thm]{Definition}
  
  \newtheorem{defn-thm}[thm]{Definition-Theorem}
   \newtheorem{ex}[thm]{Example}
   \newtheorem{example}[thm]{Example}
   \newtheorem{rem}[thm]{Remark}

   \renewcommand{\[}{\begin{equation}}
\renewcommand{\]}{\end{equation}}
\numberwithin{equation}{section}

\usepackage{amsxtra}
\usepackage{graphicx}
\usepackage{bbm,newlfont}
\usepackage{amscd}
\usepackage{hyperref}
\usepackage[german,english]{babel}
\usepackage{cancel}
\usepackage{amsmath,amsthm,amssymb}
\usepackage{enumerate}
\usepackage{color}
\usepackage{varwidth}
\usepackage{tasks}
\usepackage{mathrsfs}  
\usepackage{setspace}
\usepackage{collectbox}


\usepackage{graphicx}

\DeclareFontFamily{U}{mathx}{}
\DeclareFontShape{U}{mathx}{m}{n}{<-> mathx10}{}
\DeclareSymbolFont{mathx}{U}{mathx}{m}{n}
\DeclareMathAccent{\widehat}{0}{mathx}{"70}
\DeclareMathAccent{\widecheck}{0}{mathx}{"71}

\begin{document}
\title[Principal SUSY and nonSUSY W-algebras and their Zhu algebras]{Principal SUSY and nonSUSY W-algebras and\\ their Zhu algebras}

\author[N. Genra]{Naoki Genra$^{1}$}
\address[N. Genra]{Faculty of Science, Academic Assembly, University of Toyama 3190 Gofuku, Toyama 930--8555, Japan}
\email{genra@sci.u-toyama.ac.jp}

\author[A. Song]{Arim Song$^{2}$}
\address[A. Song]{Department of Mathematics, University of Denver, 2199 S University Blvd, Denver, CO 80210, USA}
\address{Department of Mathematical Sciences, Seoul National University, Gwanak-ro 1, Gwanak-gu, Seoul 08826, Korea}
\email{Arim.Song@du.edu}

\author[U. R. Suh]{Uhi Rinn Suh$^3$}
\address[U. R. Suh]{Department of Mathematical Sciences and Research institute of Mathematics, Seoul National University, Gwanak-ro 1, Gwanak-gu, Seoul 08826, Korea}
\email{uhrisu1@snu.ac.kr}

\thanks{$^{1}$This work was supported by the World Premier International Research Center Initiative (WPI), MEXT, Japan, and JSPS KAKENHI Grant Number JP21K20317 and JP24K16888.}
\thanks{$^{2}$This research was supported by Basic Science Research Program through the National Research Foundation of Korea(NRF) funded by the Ministry of Education(RS-2024-00409689)}
\thanks{$^{3}$This work was supported by NRF Grant \#2022R1C1C1008698 and  Creative-Pioneering Researchers Program by Seoul National University}

\begin{abstract}
This paper consists of two parts. In the first part, we prove that when $\mathfrak{g}$ is a simple basic Lie superalgebra with a principal odd nilpotent element $f$, the W-algebra $W^k(\mathfrak{g}, F)$ for $F=-\frac{1}{2}[f,f]$ is isomorphic to the SUSY W-algebra $W^k(\bar{\mathfrak{g}},f)$ via screening operators, which implies the supersymmetry of $W^k(\mathfrak{g}, F)$.  In the second part, we show that a finite SUSY W-algebra, which is a Hamiltonian reduction of $U(\widetilde{\mathfrak{g}})$ for the SUSY Takiff algebra $\widetilde{\mathfrak{g}}=\mathfrak{g}\otimes \wedge(\theta)$ is isomorphic to the Zhu algebra of a SUSY W-algebra. As a corollary, we show that a finite SUSY principal W-algebra is isomorphic to a finite principal W-algebra.
\end{abstract}

\maketitle

\section{Introduction} \label{Sec:intro}
\subsection{SUSY vertex algebras} \label{subsec:SUSY VA}
A supersymmetric (SUSY) vertex algebra $V$ introduced is a vertex algebra endowed with an odd derivation $D$ whose square is the even derivation $\partial$ of the vertex algebra. A SUSY vertex algebra corresponds to a chiral part of two-dimensional superconformal field theory in physics, which has been studied intensively since the 1980s (see, for example, \cite{EH91, FMS86, Ito91, Pope92, Wit83}). Mathematically,  Barron \cite{B1996,B2000,B2003,B2004} focused on establishing the theory in supergeometric aspects, and Heluani and Kac \cite{HK07,Kac} concentrated on constructing an algebraically oriented structural theory. They \cite{HK07} introduced the language of $\Lambda$-bracket $[ \, \cdot \, {}_\Lambda \, \cdot \, ]: V\otimes V \to \mathbb{C}[\Lambda]\otimes V$
where $\Lambda=(\lambda,\chi)$ for an odd indeterminate $\chi$ and $\lambda=-\chi^2$ as an analog of the $\lambda$-bracket of a vertex algebra \cite{BK03}. More precisely, the $\Lambda$-bracket on a SUSY vertex algebra $V$ is given by 
\begin{equation} \label{eq: intro_Lambda bracket}
        [a{}_\Lambda b]=[D a{}_\lambda b]+\chi[a{}_\lambda b], \quad a,b\in V.
\end{equation}
Here $[a{}_\lambda b]=\sum_{n\in\mathbb{Z_+}} \lambda^{(n)}a_{(n)}b$ is the $\lambda$-bracket of the vertex algebra and due to the sesquilinearity of $\Lambda$-bracket $[a{}_\Lambda b]$, the information \eqref{eq: intro_Lambda bracket} determines not only $[a{}_\lambda b]$ and $[Da{}_\lambda b]$ but also $[a{}_\lambda Db]$ and $[Da{}_\lambda Db]$. In addition, an element $\tau\in V$ such that 
\begin{equation}
    [\tau{}_\Lambda \tau] = (2\partial+3\lambda+\chi D) \tau + \frac{\lambda^2 \chi}{3} c
\end{equation}
and $D \tau$ is a conformal vector of $V$ is called a superconformal vector.

Let $\g$ be a basic Lie superalgebra with a nondegenerate supersymmetric invariant bilinear form $(\, | \, )$. The most fundamental example of a SUSY vertex algebra is the affine SUSY vertex algebra $V^k(\bar{\g})$ of level $k$. It is generated by the parity reversed vector superspace $\bar{\g}$ of $\g$ 
 and the $\Lambda$-bracket is given by 
\begin{equation} \label{eq:SUSY_affine_nosign}
    [\bar{a}{}_\Lambda \bar{b}]=(-1)^{p(a)}\big(\overline{[a,b]}+(k+h^\vee)\chi(a|b)\big)
\end{equation}
for $a,b\in \g$ and the dual Coxeter number $h^\vee$. Remarkably, the vertex subalgebra $\mathcal{F}(\g)$ generated by $\bar{\g}$ is a free field vertex algebra and $V^k(\bar{\g})\simeq V^k(\g)\otimes \mathcal{F}(\g)$ as vertex algebras. In other words, one can impose a supersymmetry on the affine vertex algebra $V^k(\g)$ by tensoring a free field vertex algebra, and the resulting SUSY vertex algebra is the SUSY affine vertex algebra $V^k(\bar{\g})$. Moreover, the SUSY vertex algebra has a superconformal vector called Kac-Todorov \cite{KT85}, and the induced conformal weights of $\bar{a}$ and $D\bar{a}$ are $\frac{1}{2}$ and $1$ for any $a\in \g$.

\subsection{SUSY W-algebras and nonSUSY W-algebras}
A W-algebra $W^k(\g)$ is introduced by Feigin-Frenkel \cite{FF90a} as a quantum Hamiltonian reduction of the affine vertex algebra $V^k(\g)$ of level $k$ associated with a Lie algebra $\g$. Afterward, Kac-Roan-Wakimoto \cite{KRW03} extended the concept and defined a W-algebra $W^k(\g, F)$ for a basic Lie superalgebra $\g$ and an even nilpotent element $F$. Its structure theory can be found in \cite{FB, DK06, KW04}, and the representation theory has also been developed (see \cite{Arakawa17} and the references in it).

On the other hand, a SUSY W-algebra was introduced in physics literature in \cite{MadRag94} and written in terms of SUSY vertex algebra in \cite{MRS21}. For an odd nilpotent element $f$ in an $\mathfrak{osp}(1|2)$ subalgebra $\mathfrak{s}$ of $\g$, a SUSY W-algebra $W^k(\bar{\g},f)$ is the quantum Hamiltonian reduction of the SUSY affine vertex algebra $V^k(\bar{\g})$ given by the SUSY analog of a BRST cohomology. In the recent papers by the second and third authors \cite{LSS23, RSS23a, RSS23b, Suh20}, various aspects of SUSY W-algebras have been studied.

Recall the relation between SUSY affine vertex algebras and affine vertex algebras described in Section \ref{subsec:SUSY VA}. In the physics literature \cite{MadRag94} by Madsen-Ragoucy, they introduced an analogous statement for W-algebras. Namely, one expects that supersymmetry can be imposed on an ordinary W-algebra by tensoring it with a suitable free-field vertex algebra. Furthermore, the resulting SUSY vertex algebra is expected to be isomorphic to a SUSY W-algebra. This paper presents a mathematical proof of this statement, in the case of W-algebras constructed from principal nilpotent elements. To be explicit, we show that the principal SUSY W-algebra is isomorphic to the corresponding ordinary W-algebra. 

To see this conjecture more carefully, we need to remind the result in \cite{DK06} that the W-algebra $W^k(\g, F)$ is strongly generated by $\text{dim}(\g^F)$ elements, where $\g^F=\Ker(\text{ad}\, F).$  On the other hand, the SUSY W-algebra $W^k(\bar{\g},f)$ is 
 strongly generated vertex algebra by $2\text{dim}(\g^f)$ elements \cite{MRS21}. From the $\mathfrak{osp}(1|2)$ representation, we have 
\begin{equation} \label{eq:F vs f}
    \g^F=\Ker(\text{ad}\, f)=\g^f\oplus [e,\g^f] 
\end{equation}
for $F=-\frac{1}{2}[f,f]$ and hence $\text{dim}(\g^F)\leq 2\cdot \text{dim}(\g^f)$, which implies the number of generators of $W^k(\g,F)$ is always less than or equal to that of $W^k(\g,f)$. 

In addition, we can compare the conformal weights of the generators for the two vertex algebras, where the conformal weights on $W^k(\g, F)$ and $W^k(\bar{\g},f)$ are induced from the conformal vector and superconformal vector, respectively. Here, the superconformal vector of $W^k(\bar{\g},f)$ is explicitly written in \cite{Song24free} 
based on the Kac-Todorov vector of $V^k(\bar{\g})$ and a superconformal vector of $bc\beta\gamma$-system \cite{SY25}. In conclusion, for any $n\in \frac{\mathbb{N}+1}{2},$ the number of weight $n$ generators for $W^k(\g,F)$ and $W^k(\bar{\g},f)$ are the same. Meanwhile, the SUSY W-algebra $W^k(\bar{\g},f)$ has $\text{dim}\g^{\mathfrak{s}}$ of weight $\frac{1}{2}$ generators where $\g^{\mathfrak{s}}$ is the centralizer of $\mathfrak{s}$ in $\g$  but every generator of the W-algebra $W^k(\g,F)$ has weight $\geq 1.$ This result can be interpreted as supporting evidence for the conjecture that the W-algebra $W^k(\g, F)$, when tensored with the free field vertex algebra $\mathcal{F}(\g^\mathfrak{s})$, possesses a supersymmetry. In particular, when $\g^\mathfrak{s}=0,$ we expect that the W-algebra $W^k(\g,F)$ itself has a supersymmetry. Note that if $f$ is an odd principal nilpotent and $F=-\frac{1}{2}[f,f]$, then $\g^\mathfrak{s}=0$. In this paper, we show $W^k(\g, F)$ indeed has a supersymmetry under these conditions on $f$ and $F$ (see Theorem \ref{thm:first thm}).

\subsection{Finite SUSY W-algebras}

As a finite counterpart of W-algebra, there is an associative algebra called finite W-algebra. Finite W-algebras were introduced by Kostant \cite{Kos78} and Premet \cite{Premet02, Premet07}  and later studied more intensively in relationship with affine W-algebras \cite{Arakawa07, DK06}. 

Let $\g$ be a semisimple Lie algebra with a $\mathbb{Z}$-grading, $\n$ be the subalgebra with positively graded elements, and $F$ be a nilpotent element with degree $-1$. The finite W-algebra $U(\g,F)$ introduced by Premet is the associative algebra 
\begin{equation} \label{eq:finite W, premet}
    (U(\g)\otimes_{U(\n)}\mathbb{C}_{-\chi})^{\textup{ad}\n}
\end{equation}
where $\chi:\n\to \mathbb{C},$ $n\mapsto (F|n)$ and $\mathbb{C}_{-\chi}$ is the 1-dimensional representation of $\n$ determined by $-\chi$. On the other hand, $U(\g, F)$ is isomorphic to the $H$-twisted Zhu algebra of the affine W-algebra $W^k(\g, F),$ where $H$ is the Hamiltonian operator induced from the conformal vector \cite{DK06}. Finite W-algebras are interesting in the context of the representation theory of affine W-algebras since the Zhu functor \cite{Zhu} explains the correspondence between positive energy simple modules of a vertex operator algebra and simple modules of its Zhu algebra \cite{FZ92}. 

A finite W-algebra for a basic Lie superalgebra $\g$ has also been investigated by various mathematicians (see, for example, \cite{ZS15, Peng21} and references therein.) Furthermore, the first author and his collaborators \cite{Genra24, GenJuil24, Gen24+} also have worked on finite W-algebras to study affine W-algebras through them. Crucially, it is still true that the finite W-algebra for a Lie superalgebra $\g$ given by \eqref{eq:finite W, premet} is isomorphic to the $H$-twisted Zhu algebra of $W^k(\g, F)$ \cite{Arakawa07, DK06}. 

In \cite{CCS24+}, the finite SUSY W-algebra  $U(\widetilde{\g},f)$ is presented as an analogue of Premet's construction \eqref{eq:finite W, premet} of a finite W-algebra $U(\widetilde{\g},f)$. More precisely, it is defined by 
\begin{equation} \label{eq:finite_SUSY_premet}
    U(\widetilde{\g},f)= ( \, U^k (\widetilde{\g})\otimes_{\widetilde{\n}}\mathbb{C}_{-\chi}\, )^{\textup{ad}\, \widetilde{\n}},
\end{equation}
 where $\widetilde{\g}=\g\otimes \wedge(\theta)$ is the SUSY Takiff algebra introduced in \cite{CCS24+}. Accordingly, the finite SUSY W-algebra arises as a Hamiltonian reduction of the SUSY Takiff algebra. It is not hard to see that the enveloping algebra of SUSY Takiff algebra is the Zhu algebra of the corresponding SUSY affine vertex algebra. Alternatively, the SUSY Takiff algebra can be considered as a supersymmetric analogue of the classical Takiff algebra originally defined in \cite{Tak71}, which is constructed as a Lie algebra over a truncated polynomial ring. Note that a Takiff superalgebra refers to a Lie superalgebra over a truncated polynomial ring in an even variable (see \cite{BR13} and references therein). In contrast, the SUSY Takiff algebra discussed in this paper is based over a ring with an odd variable.

 On the other hand, a SUSY W-algebra $W^k(\bar{\g},f)$ has a Hamiltonian operator from its superconformal vector, and hence the corresponding Zhu algebra exists. 
 This naturally leads to a comparison between the Zhu algebra of $W^k(\bar{\g},f)$ and $U(\widetilde{\g},f)$. 
 In this paper, we show these two algebras are isomorphic. In other words, the finite SUSY W-algebra is indeed a finitization of the SUSY affine W-algebra.
Therefore, in a representation theoretical view, the SUSY finite W-algebra $U(\widetilde{\g},f)$ does a crucial role in understanding the SUSY W-algebra $W^k(\bar{\g},f)$.

Unfortunately, there are not many articles that explore the representation theory of SUSY Takiff algebras or finite SUSY W-algebras beyond the results in \cite{CCS24+}. In the paper, the authors studied the category of Whittaker modules of SUSY Takiff algebras along with a Skryabin type equivalence. In the principal nilpotent cases, this category is equivalent to the category of finite SUSY W-algebra. We expect these results to be lifted to the representation theory of SUSY W-algebras.

\subsection{Outline and Main results}
Let $\g$ be a basic Lie superalgebra. In the first part of this paper, we show the isomorphism between principal W-algebras and SUSY W-algebras as vertex algebras. In Section \ref{sec: FF of W-alg}, we recall the screening operators and free field realizations of W-algebras based on the results in \cite{Genra17}. In Section \ref{sec: FF of SUSY W-alg}, we review the SUSY screening operators and free field realizations of SUSY W-algebras associated with odd principal nilpotent elements introduced in \cite{Song24free}. Section \ref{sec:W vs SUSY} focuses on the case when $f$ is an odd principal nilpotent element. Then, $F=-\frac{1}{2}[f,f]$ is automatically an even principal nilpotent. By observing the screening operators for $W^k(\g,F)$ and $W^k(\bar{\g},f)$, we obtain the following theorem.

\begin{thm} [Theorem \ref{thm:generic_isom} and Corollary \ref{cor: isom for all}] \label{thm:first thm}
    Let $f$ be a principal odd nilpotent element in $\g$. Then, the two vertex algebras are isomorphic:
    \[ W^k(\g,F)\simeq W^k(\bar{\g},f)\]
    for $F=-\frac{1}{2}[f,f]$ and all $k\neq h^\vee.$ Hence, the principal W-algebra $W^k(\g,F)$ has an $N=1$ SUSY structure.
\end{thm}

In the second part of this paper, we define finite SUSY W-algebras. In Section \ref{sec:Zhu}, we introduce the Zhu algebra $Zhu_H(W^k(\bar{\g},f))$ of a SUSY W-algebra and show that the Zhu functor commutes with the BRST cohomology (Theorem \ref{def:SUSY finite W}). In Section \ref{sec:finite W}, we define in Definition \ref{def:finite W-alg} the new associative algebra $U(\widetilde{\g},f)$ called a finite SUSY W-algebra for the SUSY Takiff algebra $\widetilde{\g}=\g\oplus \bar{\g}$ and show the following theorem.

\begin{thm} [Theorem \ref{thm:finite and Zhu}]  \label{thm:second thm}
     Let $f$ be an odd nilpotent element in an $\mathfrak{osp}(1|2)$ subalgebra of $\g$.
    Then the following two associative algebras are isomorphic:
    \[ Zhu_H(W^k(\bar{\g},f))\simeq U(\widetilde{\g},f).\]
\end{thm}

We obtain the following isomorphism as a corollary of Theorem \ref{thm:first thm} and \ref{thm:second thm}.

\begin{cor}[Theorem \ref{thm:isom_ finite}]
    Let $f$ be the odd principal nilpotent in $\g$ and $F=-\frac{1}{2}[f,f]$. Then 
    \[ U(\g,F)\simeq U(\widetilde{\g},f)\]
    where $U(\g,F)$ is the finite W-algebra associated with $\g$ and $F$.
\end{cor}

\section{Free field realization of W-algebras}\label{sec: FF of W-alg}

We recall free field realizations of the (non-SUSY) affine W-algebras. The base field is $\C$ unless otherwise noted. The superspace is a $\Z_2$-graded vector space, and elements in the homogeneous subspace of degree $\bar{0}$ (resp. $\bar{1}$) are called even (resp. odd), where $\Z_2 = \Z/2\Z \simeq \{\bar{0}, \bar{1}\}$. The $\Z_2$-degree of an element $a$ is called the parity of $a$, denoted by $p(a)$.

\subsection{Definitions}

Let $\mathfrak{g}$ be a simple basic Lie superalgebra, $(\cdot|\cdot)$ the normalized even supersymmetric invariant bilinear form on $\g$ such that $(\theta|\theta) = 2$ for an even highest root $\theta$ of $\g$, and $h^\vee$ the dual Coxeter number of $\g$. We have
\begin{align} \label{eq: bilinear form normalization}
    \kappa_\g(a|b) = 2h^\vee(a|b),\quad
    a, b \in \g,
\end{align}
where $\kappa_{\g}$ denotes the Killing form on $\g$. Choose a nilpotent element $F$ in the even part $\mathfrak{g}_{\bar{0}}$ of $\mathfrak{g}$ and a semisimple element $H$ of $\mathfrak{g}$ such that $(F, H/2)$ is a good pair. Then $\ad (H/2)$ defines a $\frac{1}{2}\Z$-grading
\begin{align}\label{eq: g grading}
\g = \bigoplus_{j \in \frac{1}{2}\Z}\g_j,\quad
\g_j = \{ a \in \g \mid [H, a]=2j a\}.
\end{align}
Let $I$ be the set of roots of $\g$. Fix a root vector $u_\alpha$ for each $\alpha \in I$. Then $I= \bigsqcup_{j \in \frac{1}{2}\Z}I_j$ with $I_j = \{ \alpha \in I \mid u_\alpha \in \g_j\}$, and there exists a set $I_{\geq}$ of positive roots such that $I_\geq \subset \bigsqcup_{j \geq 0}I_{j}$. Let $\Pi$ be the set of simple roots of $I_\geq$. Then $\Pi = \Pi_0 \sqcup \Pi_{\frac{1}{2}} \sqcup \Pi_1$, where $\Pi_j = \Pi \cap I_j$. Let $\h$ be a Cartan subalgebra of $\g$ containing $H$. Choose a basis $\{u_\alpha\}_{\alpha \in J}$ of $\h$ and extend it to a basis $\{ u_\alpha\}_{\alpha \in I \sqcup J}$ of $\g$ by adding root vectors.
\smallskip

Let $\n_\pm$ be two Lie subalgebras of $\g$ defined by
\begin{align}\label{eq: def of n_pm}
\n_+ = \bigoplus_{j >0} \g_j,\quad
\n_- = \bigoplus_{j <0} \g_j.
\end{align}
Then $\{u_\alpha\}_{\alpha \in I_+}$ is a basis of $\n_+$ with $I_+ = \bigsqcup_{j > 0}I_{j}$. Choose $u^\alpha := u_{-\alpha}$ for $\alpha \in I_+$ such that $(u^\alpha|u_\beta)=\delta_{\alpha, \beta}$. Then $\{u^\alpha\}_{\alpha \in I_+}$ is a basis of $\n_-$. For $k \in \C$, the BRST complex $C^k(\g, F)$ is defined as follows:
\begin{align*}
    C^k(\g, F) := V^k(\g) \otimes \Phi(\g_{\frac{1}{2}}) \otimes F(\mathfrak{A}),
\end{align*}
\begin{itemize}
    \item $V^k(\g)$ is the affine vertex algebra of $\g$ at level $k$, which is freely generated by basis elements of $\g$ with the $\lambda$-bracket relations
    \begin{align*}
    [a_\lambda b] = [a, b] + k(a|b),\quad
    a, b \in \g;
    \end{align*}
    \item $\Phi(\g_{\frac{1}{2}})$ is the Weyl vertex superalgebra of $\g_{\frac{1}{2}}$, which is freely generated by basis elements of
    \begin{align*}
        \Phi_{\g_{\frac{1}{2}}} \simeq \g_{\frac{1}{2}}
    \end{align*}
    with the $\lambda$-bracket relations
    \begin{align*}
    [{\Phi_{a}}_\lambda \Phi_{b}] = (F|[a, b]);
    \end{align*}
    \item $F(\mathfrak{A})$ is the charged fermion vertex superalgebra associated to $\bar{\n}_+ \oplus \bar{\n}_-$, where $\bar{\mathfrak{a}}$ denotes the parity-reversed $\mathfrak{a}$, freely generated by basis elements of
    \begin{align*}
    \varphi_{\bar{\n}_+} \simeq \bar{\n}_+,\quad
    \varphi^{\bar{\n}_-} \simeq \bar{\n}_-
    \end{align*}
    with the only non-trivial $\lambda$-bracket relations between the free generators
    \begin{align*}
        [{\varphi_{\bar{a}}}_\lambda{\varphi^{\bar{b}}}] = (a|b).
    \end{align*}
\end{itemize}
$C^k(\g, F)$ has the $\Z$-grading induced from the one on $F(\mathfrak{A})$ by $\deg\varphi^{\bar{a}} = -\deg\varphi_{\bar{b}} = 1$, $\deg\partial A = \deg A$, and $\deg\NO{AB}\ = \deg A + \deg A$ for all $A, B \in F(\mathfrak{A})$. Consider an odd element $d$ in $C^k(\g, F)$:
\begin{align*}
d =
&\sum_{\alpha \in I_{+}}(-1)^{p(\alpha)}:u_\alpha \varphi^\alpha : - \frac{1}{2}\sum_{\alpha, \beta, \gamma \in I_{+}}(-1)^{p(\alpha)p(\gamma)}c_{\alpha, \beta}^\gamma\NO{\varphi_{\gamma}\varphi^\alpha\varphi^\beta}\\
&+ \sum_{\alpha \in I_{\frac{1}{2}}}: \Phi_{\alpha}\varphi^\alpha : + \sum_{\alpha \in I_{+}}(F|u_\alpha)\varphi^\alpha,
\end{align*}
where $p(\alpha) = p(u_\alpha)$, $\Phi_{\alpha} = \Phi_{u_\alpha}$, $\varphi_\alpha = \varphi_{\bar{u}_\alpha}$, $\varphi^\alpha = \varphi^{\bar{u}^\alpha}$ and $c_{\alpha, \beta}^\gamma$ is the structure constant defined by $[u_\alpha, u_\beta]=\sum_{\gamma \in I_{+}}c_{\alpha, \beta}^\gamma u_\gamma$. It follows from \cites{FF90a, KW04} that $(C^k(\g, F)^\bullet, d_{(0)})$ forms a cochain complex. The cohomology of the complex
\begin{align*}
W^k(\g, F) := H^\bullet(C^k(\g, F), d_{(0)})
\end{align*}
is called the affine W-algebra of $\g, F$ at level $k \in \C$ \cites{FF90a, KRW}. By \cites{KW04, KW04-c}, we have $H^i(C^k(\g, F), d_{(0)}) = 0$ for $i \neq 0$.

Let $J_a$ be an element in $C^k(\g, F)$ for $a \in \g_{\leq0}$ defined by
\begin{align*}
    J_a = a + \sum_{\beta, \gamma \in I_{+}}(-1)^{p(\gamma)}c_{a, \beta}^\gamma \NO{\varphi_\gamma \varphi^\beta},
\end{align*}
where $c_{a, \beta}^\gamma$ is the structure constant determined by the formulae: $[a, u_\beta] = \sum_{\gamma \in I \sqcup J} c_{a, \beta}^\gamma u_\gamma$. Then
\begin{align}\label{eq: def of J_a and nu_k}
[J_a{}_\lambda J_b] = J_{[a, b]} 
+ \nu_k(a|b)\lambda,\quad
\nu_k(a|b) =
k(a|b)+\frac{1}{2}\kappa_\g(a|b)-\frac{1}{2}\kappa_{\g_0}(a|b),\quad
a, b \in \g_{\leq0}.
\end{align}
Hence $J_{\g_{\leq0}}$ generates a vertex subalgebra of $C^k(\g, F)$ isomorphic to the affine vertex superalgebra of $\g_{\leq0}$ at level $\nu_k$, which we denote by $V^{\nu_k}(\g_{\leq0})$. Let $C_+$ be a vertex subalgebra of $C^k(\g, F)$ generated by $J_{\g_{\leq0}}$, $\Phi_{\g_{\frac{1}{2}}}$, and $\varphi^{\bar{\n}_-}$. Then $C_+$ is $\Z_{\geq0}$-graded by the $\Z$-grading on $C^k(\g, F)$. By \cite{DK05}*{Theorem 5.7}, $(C_+^\bullet, d_{(0)})$ is a subcomplex of $C^k(\g, F)$, and
\begin{align*}
    W^k(\g, F) \simeq H^0(C_+, d_{(0)}) = \Ker\left(d_{(0)} \colon C_+^0 \rightarrow C_+^1\right) \subset C_+^0 \simeq V^{\nu_k}(\g_{\leq0}) \otimes \Phi(\g_{\frac{1}{2}}).
\end{align*}
Using a surjective vertex superalgebra homomorphism $V^{\nu_k}(\g_{\leq0}) \twoheadrightarrow V^{\nu_k}(\g_0)$ induced from $\g_{\leq0} \twoheadrightarrow \g_0$, we obtain the following map
\begin{align*}
    \mu^k \colon W^k(\g, F) \rightarrow V^{\nu_k}(\g_0)\otimes \Phi(\g_{\frac{1}{2}}),
\end{align*}
called the Miura map. The Miura map is known to be injective for all $k \in \C$ by \cites{Frenkel, Arakawa17, Nakatsuka}.

\subsection{Free field realizations} \label{subsec: FF of nonSUSY W-alg}

Recall that $\h$ is a Cartan subalgebra of $\g$ containing $H$. Let $\widehat{\h} = \h \otimes \C[t, t^{-1}] \oplus \C K$ be the affine Lie algebra of $\h$ with the Lie bracket
\begin{align*}
    [a\otimes t^m, b\otimes t^n] = m(a|b)\delta_{m+n,0}K,\quad
    [\widehat{\h}, K]=0,\quad
    a, b \in \h, \quad
    m, n \in \Z,
\end{align*}
and $\widehat{\h}_+ = \h\otimes\C[t] \oplus \C K$ a Lie subalgebra of $\widehat{\h}$. Let $\C_\alpha$ be the one-dimensional simple $\h$-module corresponding to a character $\alpha \in \h^*$. One can extend $\C_\alpha$ to an $\widehat{\h}_+$-module by $\h\otimes\C[t]t=0$ and $K=k \in \C$ on $\C_\alpha$. The induced $\widehat{\h}$-module
\begin{align*}
    \pi^k_\alpha = \operatorname{Ind}^{\widehat{\h}}_{\widehat{\h}_+}\C_\alpha := U(\widehat{\h})\underset{U(\widehat{\h}_+)}{\otimes}\C_\alpha.
\end{align*}
is called the Fock $\widehat{\h}$-module of the highest weight $\alpha \in \h^*$ at level $k$. Then $\pi^k:=\pi^k_0$ has a structure of a vertex algebra called the Heisenberg vertex algebra of $\h$ at level $k$:
\begin{align*}
    [a_\lambda b] = k(a|b)\lambda,\quad
    a, b \in \h,
\end{align*}
and $\pi^k_\alpha$ is a $\pi^k$-module, and simple if $k \neq0$.

Define an intertwining operator $\e^\alpha(z) \colon \pi^{k+h^\vee}_\beta \rightarrow \pi^{k+h^\vee}_{\alpha+\beta}\{z\}$ for $\alpha, \beta \in \h^*$ by
\begin{align*}
    \e^\alpha(z) = s_\alpha z^{\frac{\alpha_{(0)}}{k+h^\vee}}\mathrm{exp}\left(-\frac{1}{k+h^\vee}\sum_{n<0}\alpha_{(n)}\frac{z^{-n}}{n}\right)\mathrm{exp}\left(-\frac{1}{k+h^\vee}\sum_{n>0}\alpha_{(n)}\frac{z^{-n}}{n}\right),
\end{align*}
provided that $k+h^\vee \neq 0$, where $\alpha(z)=\sum_{n\in\Z}\alpha_{(n)}z^{-n-1}$ is the field on $\pi^{k+h^\vee}$ corresponding to $\alpha \in \h^* \simeq \h$, $s_\alpha$ is the shift operator mapping the highest weight vector of $\pi^k_\beta$ to that of $\pi^k_{\alpha+\beta}$ and commuting with $h_{(n)}$ for all $n \neq0$ and $h \in \h$. We have
\begin{align*}
    h(z)\e^{\alpha}(w) \sim \frac{\alpha(h)\e^{\alpha}(w)}{z-w},\quad
    \partial_z \e^{\alpha}(z) = \frac{1}{k+h^\vee}\NO{\alpha(z)\e^{\alpha}(z)}.
\end{align*}

Suppose that $\g_0=\h$. Then notice that 
\begin{align*}
    \nu_k(a|b) = (k+h^\vee)(a|b),\quad
    a, b \in \h.
\end{align*}
Hence, $V^{\nu_k}(\g_0) = \pi^{k+h^\vee}$ and the Miura map is
\begin{align*}
    \mu^k \colon W^k(\g, F) \hookrightarrow  \pi^{k+h^\vee} \otimes \Phi(\g_{\frac{1}{2}}).
\end{align*}
\begin{thm}[\cite{Genra17}]\label{thm:non-SUSY_W}
    Suppose that $\g_0=\h$ and that $k$ is generic. Then
    \begin{align*}
        W^k(\g, F) \simeq \Img \mu^k = \bigcap_{\alpha \in \Pi}\Ker\left(\int S_\alpha(z)\,dz \colon \pi^{k+h^\vee} \otimes \Phi(\g_{\frac{1}{2}}) \rightarrow \pi^{k+h^\vee}_{-\alpha}\otimes \Phi(\g_{\frac{1}{2}}) \right),
    \end{align*}
    where $S_\alpha(z) \colon \Phi(\g_{\frac{1}{2}}) \otimes \pi^{k+h^\vee} \rightarrow \Phi(\g_{\frac{1}{2}}) \otimes \pi^{k+h^\vee}_{-\alpha}\{z\}$ is an intertwining operator defined by
    \begin{align*}
        S_\alpha(z) =
        \begin{cases}
           \e^{-\alpha}(z) & \mathrm{if}\ \alpha \in \Pi_1,\\
            \Phi_{\alpha}(z)\e^{-\alpha}(z) & \mathrm{if}\ \alpha \in \Pi_{\frac{1}{2}}.
        \end{cases}
    \end{align*}
\end{thm}
The operators $\displaystyle \int S_\alpha(z)\,dz := \underset{z=0}{\mathrm{Res}}\,S_\alpha(z)$ are called the \textit{screening operators} of $W^k(\g, F)$. Theorem \ref{thm:non-SUSY_W} gives a free field realization of $W^k(\g, F)$ in terms of the screening operators.

\begin{rem}
    In the original paper \cite{Genra17}, the condition
    \begin{align*}
        (F|u_\alpha) \neq 0,\quad
        \alpha \in \Pi_1
    \end{align*}
    is assumed for the screening operator $\displaystyle \int S_\alpha(z)\,dz$ to exist, but it is always true.
\begin{proof}
Write $F$ as a linear combination of (negative) root vectors in $\g_{-1}$: $F=\sum_{\beta \in I_1}c_\beta u_{-\beta}$. If $(F|u_\alpha) = 0$ for some $\alpha \in \Pi_1$, then $c_\alpha =0$. Using the fact that $\ad F \colon \g_1 \rightarrow \g_0$ is injective, it follows that $[F, u_\alpha] = \sum_{\beta \in I_1\backslash\{\alpha\}} c_\beta [u_{-\beta}, u_\alpha]$ is non-zero. Thus, there exists $\gamma=-\beta+\alpha \in I$ for some $\beta \in I_1 \backslash\{\alpha\}$ such that $u_\gamma \in \g_0$, contrary to our assumption that $\g_0 = \h$.
\end{proof}
\end{rem}

\section{Free field realization of SUSY W-algebras} \label{sec: FF of SUSY W-alg}
We recall the definition of supersymmetric(SUSY) W-algebras and introduce their free field realization. As in Section \ref{sec: FF of W-alg}, we set $\g$ to be a finite-dimensional simple basic classical Lie superalgebra with the invariant bilinear form $(\cdot |\cdot)$ such that $(\theta|\theta)=2$ for an even highest root $\theta$ of $\g$. Assume that $\g$ has a subalgebra $\mathfrak{s}$ isomorphic to $\mathfrak{osp}(1|2)=\textup{Span}_{\C}\{E,e,H,f,F\}$, whose even subspace $(E,H,F)$ forms an $\mathfrak{sl}_2$ triple.

\subsection{SUSY W-algebras} \label{subsec: SUSY W-alg}
With respect to the adjoint action of $\frac{H}{2}$, the Lie superalgebra $\g$ decomposes into the direct sum of eigenspaces \eqref{eq: g grading}. Let $\n_+$ and $\n_-$ be the two vector subspaces defined by \eqref{eq: def of n_pm}. For $k\in \C$, the SUSY BRST complex $C^k(\bar{\g},f)$ is defined as follows:

\begin{equation}
    C^k(\bar{\g},f):=V^k(\bar{\g})\otimes F(\bar{\mathfrak{A}}),
\end{equation}
\begin{itemize}
    \item $V^k(\bar{\g})$ is the SUSY affine vertex algebra of level $k$, i.e., it is the SUSY vertex algebra freely generated by the basis elements of $\bar{\g}$, the parity-reversed $\g$, whose $\Lambda$-bracket relation between $a, b\in \g$ are given by
    \begin{equation} \label{eq: susy affine bracket}
        [\bar{a}{}_{\Lambda}\bar{b}]=(-1)^{p(a)(p(b)+1)}\overline{[a,b]}+\chi(a|b)(k+h^{\vee});
    \end{equation}
    \item $F(\bar{\mathfrak{A}})$ is the SUSY charged free fermion vertex algebra associated with $\g$, i.e., the SUSY vertex algebra $F(\bar{\mathfrak{A}})$ is freely generated by the basis elements of
    \begin{equation*}
        \phi^{\bar{\n}_-}\simeq \bar{\n}_-, \quad \phi_{\n_+}\simeq \n_+,
    \end{equation*}
    where the only nontrivial relation between the free generators is
    \begin{equation}
        [\phi^{\bar{a}}{}_{\Lambda}\phi_{b}]=(a|b).
    \end{equation}
\end{itemize}

\begin{rem} \label{rem:isom of affines_sect3}
The $\Lambda$-bracket \eqref{eq: susy affine bracket} is the same as the bracket \eqref{eq:SUSY_affine_nosign} up to the sign $(-1)^{p(a)p(b)}.$ These two $\Lambda$-brackets define isomorphic vertex algebras via the map given by 
\[ \bar{a}\mapsto \sqrt{-1}^{p(a)}\bar{a}, \quad  D\bar{a}\mapsto \sqrt{-1}^{p(a)}D\bar{a}.\]
\end{rem}

Consider the following even element in $C^k(\bar{\g},f)$
\begin{align}
  d=\sum_{\alpha\in I_+} :\bar{u}_{\alpha} \phi^{\alpha}:+\frac{1}{2}\sum_{\alpha,\beta\in I_+}(-1)^{p(\alpha)(p(\beta)+1)}:\!\phi_{[u_{\alpha},u_{\beta}]}\phi^{\beta}\phi^{\alpha}\!:-\sum_{\alpha\in I_+} (f|u_{\alpha})\phi^{\alpha},
\end{align}
where $\phi^{\alpha}=\phi^{\bar{u}^{\alpha}}$ and $\phi_{\alpha}=\phi_{u_{\alpha}}$ for each $\alpha\in I_+$. It is shown in \cite{MRS21} that the odd endomorphism $d_{(0|0)}$ is an odd derivation on $C^k(\bar{\g}, f)$ whose square vanishes, where 
\begin{equation}
    d_{(0|0)}(A):=[\, d \, {}_{\Lambda}\, A \, ] \, |_{\chi=\lambda=0}, \quad A\in C^k(\bar{\g},f).
\end{equation}
Therefore, one can define the cohomology with this derivation.
\begin{defn}[\cite{MRS21}] \label{def: susy W-alg}
  Let $\g$ be a finite basic simple Lie superalgebra with an odd nilpotent $f$ that lies in a subalgebra isomorphic to $\mathfrak{osp}(1|2)$.
  The \textit{SUSY W-algebra associated with $\g$ and $f$ of level $k\in \C$} is defined by
  \begin{equation*}
    W^{k}(\bar{\g},f):=H(C^{k}(\bar{\g},f), d_{(0|0)}).
  \end{equation*}
\end{defn}

For each $a\in \g$, let $J_{\bar{a}}$ be an element in $C^k(\bar{\g},f)$ defined by
\begin{equation} \label{eq: building block SUSY}
    J_{\bar{a}}=\bar{a}+\sum_{\beta,\gamma\in I_+}(-1)^{(p(a)+1)(p(\beta)+1)}(u^{\gamma}|[u_{\beta},a]):\!\phi^{\beta} \phi_{\gamma}\!:.
  \end{equation}
From the relation
\begin{equation} \label{eq: building block bracket}
    [J_{\bar{a}}{}_{\Lambda}J_{\bar{b}}]=(-1)^{p(a)(p(b)+1)}J_{\overline{[a,b]}}+\chi (a|b)(k+h^{\vee})
\end{equation}
for $a,b\in \g_{\leq 0}$, one can see that $J_{\bar{\g}_{\leq 0}}$ generates a SUSY affine vertex algebra. We denote this subalgebra by $V^{\zeta_k}(\bar{\g}_{\leq 0})$. The level $\zeta_k$ of this algebra is $k$ shifted by $\frac{1}{2}\kappa_{\g}-\frac{1}{2}\kappa_{\g_{\leq 0}}$, as one can see in Remark \ref{rem: susy affine level}. Consider the subalgebra of the SUSY BRST complex generated by $J_{{\bar{\g}}_{\leq 0}}\oplus \phi^{\bar{\n}_-}$ and denote it by $\widetilde{C}^k(\bar{\g}, f)$. As a result of the similar argument for usual W-algebras, $(\widetilde{C}^k(\bar{\g},f),d_{(0|0)})$ forms a cochain complex whose cohomology is concentrated on the charge $0$ space and satisfies
\begin{equation*}
    \begin{aligned}
    W^k(\bar{\g},f)&\simeq H(\widetilde{C}^k(\bar{\g},f), d_{(0|0)})=H^0(\widetilde{C}^k(\bar{\g},f), d_{(0|0)})\\
    &=\Ker\big(d_{(0|0)}:(\widetilde{C}^k)^0\rightarrow (\widetilde{C}^k)^1\big)\subset(\widetilde{C}^k)^0\simeq V^{\zeta_k}(\bar{\g}_{\leq 0}),
    \end{aligned}
\end{equation*}
where $(\widetilde{C}^k)^0$ and $(\widetilde{C}^k)^1$ are the charge $0$ and charge $1$ subspaces of $\widetilde{C}^k(\bar{\g},f)$, respectively.
\begin{rem}[\cite{Song24free}] \label{rem: susy affine level}
    For any finite-dimensional Lie superalgebra $\g$ with a supersymmetric bilinear form $(\cdot |\cdot)$, the SUSY affine vertex algebra of level $k$ is defined similarly to \eqref{eq: susy affine bracket} but with the $\Lambda$-bracket replaced by
    \begin{equation} \label{eq: susy affine gen bracket}
        [\bar{a}{}_{\Lambda}\bar{b}]=(-1)^{p(a)(p(b)+1)}\overline{[a,b]}+\chi \tau_k(a|b), \quad  \tau_k(a|b)=k(a|b)+\frac{1}{2}\kappa_{\g}(a|b).
    \end{equation}
    Since we have the relation \eqref{eq: bilinear form normalization} for a simple or abelian Lie superalgebra $\g$, \eqref{eq: susy affine gen bracket} coincides with \eqref{eq: susy affine bracket} in these cases. Also, note that if we replace $\tau_k$ by the bilinear form $\tau_k+\frac{1}{2}\kappa_{\g}-\frac{1}{2}\kappa_{\g_{\leq0}}$ defined on $\g_{\leq 0}$, it recovers the $\Lambda$-bracket relation in \eqref{eq: building block bracket}. Therefore, we use the notation $V^{\zeta_k}(\bar{\g}_{\leq 0})$ for $\zeta_k:=k+\frac{1}{2}\kappa_{\g}-\frac{1}{2}\kappa_{\g_{\leq 0}}$ to denote the SUSY vertex algebra generated by $J_{\bar{a}}$'s for $a\in \g_{\leq 0}$.
\end{rem}

\subsection{Supersymmetric Miura map}
Recall from Section \ref{subsec: SUSY W-alg} that the SUSY W-algebra $W^k(\bar{\g},f)$ can be realized as a vertex subalgebra of $V^{\zeta_k}(\bar{\g}_{\leq 0})$. This SUSY affine vertex algebra has a natural projection map $V^{\zeta_k}(\bar{\g}_{\leq 0}) \twoheadrightarrow V^{\nu_k}(\bar{\g}_0)$, where $V^{\nu_k}(\bar{\g}_0)$ is the SUSY vertex subalgebra generated by $J_{\bar{a}}$'s for $a\in \g_0$. To align with the notation introduced in Remark \ref{rem: susy affine level}, we denote the level by $\nu_k$ defined in \eqref{eq: def of J_a and nu_k}. With the use of the projection, one obtains the \textit{supersymmetric Miura map}
\begin{equation} \label{eq: susy miura}
    \bar{\mu}^k\colon W^k(\bar{\g},f)\rightarrow V^{\nu_k}(\bar{\g}_0).
\end{equation}

Note that $W^k(\bar{\g},f)$ and $V^{\nu_k}(\bar{\g}_0)$ have conformal $\frac{1}{2}\Z_+$-gradings for any $k\neq -h^{\vee}$, and the map \eqref{eq: susy miura} preserves the conformal grading. It easily follows from \cites{Song24free,KT85} that the SUSY W-algebra has a conformal grading given by $\Delta(J_{\bar{a}})=\frac{1}{2}-j$ for $a\in \g_{j}$, while $\Delta(\bar{a})=\frac{1}{2}$ for each $a\in \g_0$ in $V^{\nu_k}(\bar{\g}_0)$. Applying the argument in \cite{Nakatsuka}, we get the injectivity of $\bar{\mu}^k$ for $k\neq -h^{\vee}$.

\begin{prop} \label{prop: miura injectivity}
    For $k\neq -h^{\vee}$, the supersymmetric Miura map $\bar{\mu}^k$ is injective.
\end{prop}
\begin{proof}
The proof is a direct analogue of that given in \cite{Nakatsuka} for W-superalgebras.

By \cite[Lemma 15]{Nakatsuka}, the injectivity of $\bar{\mu}^k$ follows from the injectivity of 
\begin{equation} \label{eq: graded miura}
    \textup{gr} \bar{\mu}^k\colon \textup{gr} W^k(\bar{\g},f)\rightarrow \textup{gr} V^{\nu_k}(\bar{\g}_0),
\end{equation}
which is the induced map on the graded spaces with respect to the Li's filtration. Recall from \cite{Suh20} that the graded space $\textup{gr}W^k(\bar{\g},f)$ can be identified with the invariant space under the adjoint action of $\g_+:=\g_{>0}$ and its super partner, by using the subcomplex $\widetilde{C}(\bar{\g},f)$ generated by $J_{\bar{\g}_{\leq 0}}\oplus \phi^{\bar{n}_-}$.

Denote the Zhu's $C_2$-algebra of the graded space and the induced differential by $(\textup{gr}^{\textup{fin}}\widetilde{C}^k(\bar{\g},f),Q^{\textup{fin}})$. Following the proof in \cite{Suh20} along with the finitization, it can be easily shown that $\textup{gr}^{\textup{fin}}\widetilde{C}^k(\bar{\g},f)\simeq S(\widetilde{\g}_{\leq 0})\otimes \wedge \widetilde{\g}_+^*$ is the Chevalley-Eilenberg complex of $\widetilde{\g}_+$ with coefficients in $S(\widetilde{\g}_{\leq 0})$, where $\widetilde{\g}=\g\oplus \bar{\g}$ is the Lie superalgebra with the Lie bracket \eqref{eq: Lie bracket Takiff}. Note that the left $\widetilde{\g}_+$-module structure of $S(\widetilde{\g}_{\leq 0})$ is isomorphic to the coordinate ring $\C[f+\widetilde{\g}_{\geq 0}]$, whose action is determined by the right adjoint action of $\widetilde{G}_+$, the unipotent algebraic supergroup whose Lie superalgebra is $\widetilde{\g}_+$. 

For each $p\in \Z/2$, denote $\widetilde{\g}_p:= \g_p\oplus \bar{\g}_p$. Let $\widetilde{S}_f:=f+\widetilde{\g}^e$, where $\widetilde{\g}^e \subset \widetilde{\g}$ is the kernel of the adjoint action of $e$.
Since each eigenspace is decomposed as $\widetilde{\g}_p=\widetilde{\g}_p^e\oplus[f,\widetilde{\g}_{p+\frac{1}{2}}]$ for $\widetilde{\g}_p^e=\widetilde{\g}^e \cap \widetilde{\g}_p$, one can show that
\begin{equation*}
    \widetilde{S}_f\times \widetilde{G}_+ \simeq f+\widetilde{\g}_{\geq 0}
\end{equation*}
following the computations in \cite[Proposition 18]{Nakatsuka}. As a result, the finitization of \eqref{eq: graded miura} can be identified with 
\begin{equation*}
    \bar{\mu}^{\textup{fin}}: \C[\widetilde{S}_f]\simeq \C[f+\widetilde{\g}_{\geq 0}]^{\widetilde{G}_+}\rightarrow \C[f+\widetilde{\g}_0].
\end{equation*}
The injectivity of $\bar{\mu}^{\textup{fin}}$ can be checked by showing that the image of
\begin{equation}
\widetilde{G}_+\times (f+\widetilde{\g}_0)\rightarrow f+\widetilde{\g}_{\geq 0} \twoheadrightarrow f+\widetilde{\g}^e=\widetilde{S}_f
\end{equation}
is Zariski dense. For each $X\in\widetilde{\g}_0$, considering $\exp(e)\cdot (f-\frac{1}{3}H+X)$ shows the desired statement. Finally, the injectivity of \eqref{eq: graded miura} follows, since $W^k(\bar{\g},f)$ is freely generated by the fields associated with $\widetilde{\g}^e\simeq f+\widetilde{\g}^e=\widetilde{S}_f$.
\end{proof}

\subsection{Free field realizations}

Let $\h$ be a Cartan subalgebra of $\g$ containing $H$ as in Section \ref{sec: FF of W-alg}. Let us denote by $\bar{\h}$ the parity-reversed space of $\h$. Then the \textit{SUSY Heisenberg algebra $\mathcal{H}^{\mathfrak{h}}$ associated with $\mathfrak{h}$} is the infinite-dimensional Lie superalgebra
\begin{equation*}
    \mathcal{H}^{\mathfrak{h}}:=\left(\bar{\h}\oplus \h\right)\otimes \C[t, t^{-1}]\oplus \C K
\end{equation*}
whose Lie bracket relations among generators are as follows:
\begin{equation}
\begin{gathered} \label{eq: susy Heisenberg}
  [a\otimes t^m, b\otimes t^n]=m(a|b)\delta_{m+n,0}K,\\
  [\bar{a}\otimes t^m,\bar{b}\otimes t^n]=(a|b)\delta_{m+n,-1}K,\quad  [a\otimes t^m,  \bar{b}\otimes t^n]=0
\end{gathered}
\end{equation}
  for $a,b\in \h$. Note here that $a\otimes t^m$'s and $K$ are even with central $K$, while $\bar{b}\otimes t^n$'s are odd. One can easily observe from \eqref{eq: susy Heisenberg} that $\mathcal{H}^{\h}$ contains the affine Lie algebra $\widehat{\h}$. Now, considering a Lie subalgebra
  \begin{equation} \label{eq: borel susy Heisenberg}
    \mathcal{H}^{\h}_+:=(\bar{\h}\oplus\h)\otimes \C[t]\oplus \C K
  \end{equation}
  as a Borel subalgebra, one can define the Fock modules for the SUSY Heisenberg algebras.
  \begin{defn}
    For each $\alpha\in \h^*$ and $k \in \C$, let $\C_{\alpha}:=\C|\alpha\rangle$ (resp., $\overline{\C}_{\alpha}:=\C|\alpha\rangle$) be the $\mathcal{H}^{\h}_+$-module where 
    \begin{equation*}
        h|\alpha\rangle=\alpha(h)|\alpha\rangle,\quad  h\in \h,
    \end{equation*}
    while $\bar{\h}\otimes \C[t]\oplus \h \otimes \C[t]t$ acts as zero on $|\alpha\rangle$, $K=k$ and $|\alpha\rangle$ is even (resp., odd). Then, we define the \textup{even Fock module} and the \textup{odd Fock module} of $\mathcal{H}^{\mathfrak{h}}$ of highest weight $\alpha\in \h^*$ as follows:
    \begin{equation*}
        \bar{\pi}_{\alpha}^{k}=\textup{Ind}_{\mathcal{H}^{\h}_+}^{\mathcal{H}^{\h}}\C_{\alpha}, \quad P\bar{\pi}_{\alpha}^{k}=\textup{Ind}_{\mathcal{H}^{\h}_+}^{\mathcal{H}^{\h}}\overline{\C}_{\alpha},
    \end{equation*}
    where $P$ denotes the parity change.
  \end{defn}
Especially, the even Fock module $\bar{\pi}^{k+h^\vee}:=\bar{\pi}^{k+h^\vee}_0$ has a SUSY vertex algebra structure isomorphic to $V^k(\bar{\h})$ under the isomorphism $\bar{\pi}^{k+h^\vee}\rightarrow V^k(\bar{\h})$
\begin{equation*}
    \begin{aligned}
    (h_{i_1}\otimes t^{-m_1})&\cdots (h_{i_r}\otimes t^{-m_r})(\bar{h}_{j_1}\otimes t^{-n_1})\cdots(\bar{h}_{j_s}\otimes t^{-n_s})\\
    &\mapsto (D\bar{h}_{i_1})_{(-m_1)}\cdots (D\bar{h}_{i_r})_{(-m_r)}(\bar{h}_{j_1})_{(-n_1)}\cdots (\bar{h}_{j_s})_{(-n_s)}|0\rangle
    \end{aligned}
\end{equation*}
for $h_i\in \h$ and positive integers $m_i$'s and $n_j$'s. Thus, we regard $\bar{\pi}^{k+h^\vee}$ as a SUSY vertex algebra, which we call the \textit{SUSY Heisenberg vertex algebra associated with $\h$ at level $k$}.

For each simple root $\alpha$ of $\g$ and $k+h^\vee \neq0$, we use the notations as follows:
\begin{equation}
    \widehat{\pi}^{k+h^\vee}:=\bar{\pi}^{k+h^\vee}_0, \quad \quad
    \widehat{\pi}^{k+h^\vee}_{-\alpha}:=
    \left\{
    \begin{array}{ll}
    \bar{\pi}^{k+h^\vee}_{-\alpha} & \textup{if }\alpha\textup{ is odd,}\\
    P\bar{\pi}^{k+h^\vee}_{-\alpha} & \textup{if }\alpha\textup{ is even.}
    \end{array}
    \right.
\end{equation}

Next, we use the formal Laurent series in variable $Z=(z,\theta)$, where $z$ is even, and $\theta$ is an odd variable with $\theta^2=0$. We refer to \cite{HK07} for the calculus of formal series in $Z$. For each simple root $\alpha\in \Pi$, define $\e^{-\alpha}(Z)$ to be the formal Laurent series in $\textup{End}_{\C}(\widehat{\pi}^{k+h^\vee}, \widehat{\pi}_{-\alpha}^{k+h^\vee})$ by
\begin{equation} \label{eq: exponential definition}
    \begin{aligned}
      & \e^{-\alpha}(Z)\\
      & =
      s_{-\alpha}\exp\Big(\!-\frac{1}{k+h^\vee}\sum_{n<0}Z^{-n-1|1}\alpha_{(n|1)}\Big)\exp\Big(\,\frac{1}{k+h^\vee}\sum_{n<0}\frac{Z^{-n|0}}{n}\alpha_{(n|0)}\Big)\\
      &\ \exp\Big(\,\frac{1}{k+h^\vee}\sum_{n>0}\frac{Z^{-n|0}}{n}\alpha_{(n|0)}\Big)\exp\Big(\!-\frac{1}{k+h^\vee}\sum_{n\geq 0}Z^{-n-1|1}\alpha_{(n|1)}\Big)Z^{-\frac{\alpha_{(0|0)}}{k+h^\vee}|0}.
    \end{aligned}
  \end{equation}
  In \eqref{eq: exponential definition}, the exponential of the series is defined via the power sum, and the element $\alpha\in \widehat{\pi}^{k+h^\vee}$ is given by $\alpha=(\bar{h}_{\alpha})_{(-1)}\vac \in \widehat{\pi}^{k+h^\vee}$ for the coroot $h_{\alpha}\in \h$ of $\alpha$. Also, $s_{-\alpha}:\widehat{\pi}^{k+h^\vee}\rightarrow \widehat{\pi}_{-\alpha}^{k+h^\vee}$ is the linear operator with the opposite parity of the root $\alpha$, which is determined by the properties
\begin{equation} \label{eq: commutator with shift operator}
  \begin{aligned}
  & s_{-\alpha}\vac=\left|-\alpha\right>, \\
  & [\bar{h}_{(n|i)},s_{-\alpha}]=0 \text{ for } (n|i)\neq 0, \\
  & [\bar{h}_{(0|0)}, s_{-\alpha}]=-\frac{\alpha(h)s_{-\alpha}}{k+h^\vee}
  \end{aligned}
\end{equation} 
for any $h\in \h$. 

Assume $\g_0=\h$. Then, the target space $V^{\nu_k}(\bar{\g}_0)$ of $\bar{\mu}^k$ is isomorphic to the SUSY Heisenberg vertex algebra $\widehat{\pi}^{k+h^\vee}$. Therefore, one can rewrite \eqref{eq: susy miura} as
\begin{equation*}
    \bar{\mu}^k: W^k(\bar{\g},f) \rightarrow \widehat{\pi}^{k+h^\vee}.
\end{equation*}

\begin{thm}[\cite{Song24free}] \label{thm:susy_W}
    Assume that $\g_0=\h$. Then for generic $k\in \C$,
    \begin{equation*}
        W^k(\bar{\g},f)\simeq \operatorname{Im}\bar{\mu}^k=\bigcap_{\alpha\in \Pi}\Ker\Big(\int \e^{-\alpha}(Z)\,dZ : \widehat{\pi}^{k+h^\vee}\rightarrow \widehat{\pi}_{-\alpha}^{k+h^\vee}\Big)
    \end{equation*}
    where $\int e^{-\alpha}(Z)\,dZ$ is the super residue of the series $\e^{-\alpha}(Z)$.
\end{thm}

\section{W-algebras vs. SUSY W-algebras} \label{sec:W vs SUSY}

Assume that $\g$ is a simple basic Lie superalgebra that admits a principal odd nilpotent $f$. In other words, $\g$ has an $\mathfrak{osp}(1|2)=\textup{Span}_{\C}\{E,e,H,f,F\}$ embedding satisfying $F=-\frac{1}{2}[f,f]$ and $\textup{dim}(\g^f)=\textup{rank}(\g)$ for $\Ker(\textup{ad}f)=\g^f$. In this section, we show that the W-algebra $W^k(\g,F)$ and SUSY W-algebra $W^k(\bar{\g},f)$ are isomorphic for any $k\neq -h^{\vee}$.

Notice first that the even nilpotent $F$ is also principal in $\g$. Considering the adjoint action of the principal $\mathfrak{osp}(1|2)$ subalgebra, the Lie superalgebra $\g$ is decomposed as a sum of irreducible $\mathfrak{osp}(1|2)$-modules as
\begin{equation*}
    \g=R_1\oplus \cdots \oplus R_l,
\end{equation*}
where $l=\textup{rank}(\g)$ and each $R_i$ has dimension$\geq 3$ whose weight spaces are $1$-dimensional with homogeneous parity \cite{FRS93}. Therefore, one can form a basis for $\g_{\bar{0}}^F=\g_{\bar{0}}\cap \Ker(\textup{ad}F)$ by taking the even nonzero vectors of lowest weight in $R_i$'s. It implies that $\dim (\g_{\bar{0}}^F)=l=\textup{rank}(\g)$, that is, $F$ is also principal in $\g$.

\begin{thm} \label{thm:generic_isom}
    For generic $k$, the principal SUSY W-algebra of level $k$ is isomorphic to the principal W-algebra of level $k$.
\end{thm}
\begin{proof}
Take the set of simple roots $\Pi=\{\alpha_1, \cdots,\alpha_l\}$ to be purely odd satisfying $\Pi=\Pi_{\frac{1}{2}}$. Then, one can choose the odd principal nilpotent $f$ as
\begin{equation*}
    f=e^{\alpha_1}+\cdots +e^{\alpha_l},
\end{equation*}
where $\{e_{\alpha}\}_{\alpha\in I_+}$ and $\{e^{\alpha}\}_{\alpha\in I_+}$ are the dual bases of $\g_{>0}$ and $\g_{<0}$ satisfying $(e_{\alpha}|e^{\beta})=\delta_{\alpha, \beta}$.

We show that the image of $W^k(\g, F)$ in Theorem \ref{thm:non-SUSY_W} coincides with the image of $W^k(\bar{\g},f)$ in Theorem \ref{thm:susy_W} by comparing the screening operators. First, the domains of the screening operators are isomorphic due to the following vertex algebra homomorphism
\begin{equation} \label{eq:domain_screening}
    \tau\colon \pi^{k+h^\vee}\otimes\Phi(\g_{\frac{1}{2}}) \DistTo \widehat{\pi}^{k+h^\vee}, \quad \quad \sqrt{k+h^\vee}\Phi_{\alpha}\mapsto \overline{h}_{\alpha},\quad h \mapsto D\bar{h},
\end{equation}
where $h_{\alpha}\in \h$ is the coroot of the simple root $\alpha$. It follows from the fact that $h_{\alpha}=[f,e_{\alpha}]$ and
\begin{equation*}
    [\Phi_{\alpha}{}_{\lambda}\Phi_{\beta}]=(F|[e_{\alpha},e_{\beta}])=([f,e_{\alpha}]|[f,e_{\beta}])
\end{equation*}
for any simple roots $\alpha$ and $\beta$. Using the isomorphism $\tau$, one can consider the SUSY Fock module $\widehat{\pi}_{-\alpha}^{k+h^\vee}$ as a $\pi^{k+h^\vee}\otimes\Phi(\g_{\frac{1}{2}})$-module. For the highest weight vector $|\!-\alpha\rangle\in \widehat{\pi}_{-\alpha}^{k+h^\vee}$, we have $\tau(h)_{(0)}|\!-\alpha\rangle=(D\bar{h})_{(0)}|\!-\alpha\rangle=-\alpha(h)|\!-\alpha\rangle$. Therefore, as $\pi^{k+h^\vee}\otimes\Phi(\g_{\frac{1}{2}})$-modules,
\begin{equation} \label{eq:codomain_screening}
     \pi_{-\alpha}^{k+h^\vee}\otimes\Phi(\g_{\frac{1}{2}})\simeq
    \widehat{\pi}_{-\alpha}^{k+h^\vee},
\end{equation}
which gives an isomorphism of the codomains of the screening operators.

Now, Theorem \ref{thm:non-SUSY_W} and Theorem \ref{thm:susy_W} with $\Pi=\Pi_{\frac{1}{2}}$ show that if $k$ is generic, the Miura map and the supersymmetric Miura map induce isomorphisms
\begin{align*}
    &W^k(\g, F) \simeq \bigcap_{\alpha \in \Pi}\Ker\int\Phi_{\alpha}(z)\e^{-\alpha}(z)\,dz,\\
    &W^k(\bar{\g}, f) \simeq \bigcap_{\alpha \in \Pi}\Ker\int\e^{-\alpha}(Z)\,dZ.
\end{align*}
Under the identifications \eqref{eq:domain_screening} and \eqref{eq:codomain_screening}, we have
\begin{align*}
    \int\Phi_{\alpha}(z)\e^{-\alpha}(z)\,dz = \int\e^{-\alpha}(Z)\,dZ
\end{align*}
for all $\alpha \in \Pi$. Therefore, $W^k(\g, F) \simeq W^k(\bar{\g}, f)$ for generic $k$.
\end{proof}

We introduce a \textit{continuous family of vector spaces} as in \cite{CGN} (the same idea was also used in \cite{AFO}). Let $\{V^k\}_{k\in\C}$ be a family of vector subspaces of the same dimension $d$ in the fixed vector space $W$. $\{V^k\}_{k\in\C}$ is called continuous if the induced map $\C \ni k \mapsto V^k \in \mathrm{Gr}(d, W)$ is continuous, where $\mathrm{Gr}(d, W)$ is the Grassmannian manifold.
\smallskip

Suppose that $k \neq - h^\vee$. Let $\pi:= \pi^1$ and $\widehat{\pi} := \widehat{\pi}^1$. Then $\pi^{k+h^\vee} \simeq \pi$ and $\widehat{\pi}^{k+h^\vee} \simeq \widehat{\pi}$ by $h/\sqrt{k+h^\vee} \mapsto h$ for all $h\in\h$. Since the principal SUSY/non-SUSY W-algebras $W^k(\bar{\g}, f)$, $W^k(\g, F)$ and the SUSY Heisenberg vertex algebras $\widehat{\pi} \simeq \Phi(\g_{\frac{1}{2}})\otimes\pi$ are $\frac{1}{2}\Z_{\geq0}$-graded by the conformal weights, the Miura maps induce injective maps between finite-dimensional vector spaces for each $k$ and $\Delta \in \frac{1}{2}\Z_{\geq0}$:
    \begin{align*}
        W^k(\bar{\g}, f)_\Delta\quad \hookrightarrow \quad\widehat{\pi}_\Delta\quad
        \hookleftarrow \quad W^k(\g, F)_\Delta,
    \end{align*}
where $V_\Delta$ denotes the homogeneous subspace of $V$ of the conformal weight $\Delta$. Then $W^k(\bar{\g}, f)_\Delta$ and $W^k(\g, F)_\Delta$ can be thought of subspaces of the fixed vector spaces $\widehat{\pi}_\Delta$. Thus, $\displaystyle \left\{W^k(\bar{\g}, f)_\Delta\right\}_{k \in \C\backslash\{-h^\vee\}}$ and $\displaystyle \left\{W^k(\g, F)_\Delta\right\}_{k \in \C\backslash\{-h^\vee\}}$ define continuous families of vector spaces.

\begin{cor} \label{cor: isom for all}
    Theorem \ref{thm:generic_isom} holds for all $k \in \C\backslash\{-h^\vee\}$.
    \begin{proof}
    Apply \cite[Lemma 5.14]{CGN} with Theorem \ref{thm:generic_isom} for the continuous families
    \begin{align*}
    \left\{W^k(\bar{\g}, f)_\Delta\right\}_{k \in \C\backslash\{-h^\vee\}}\text{ and } \left\{W^k(\g, F)_\Delta\right\}_{k \in \C\backslash\{-h^\vee\}}
    \end{align*}
    of vertex superalgebras.
    \end{proof}
\end{cor}

By the classification \cite{LSS86,FRS93} of Lie superalgebras $\g$ admitting a principal embedding of $\mathfrak{osp}(1|2)$, Corollary \ref{cor: isom for all} applies when $\g$ is one of the following:
\begin{equation} \label{eq: classification}
    \mathfrak{sl}(n\pm 1|n),\, \mathfrak{osp}(2n\pm 1|2n),\, \mathfrak{osp}(2n+2|2n),\, \mathfrak{osp}(2n|2n),\, D(2,1;\alpha) \textup{ with } \alpha\in \C\setminus\{0, \pm 1\}.
\end{equation}
It is important to note that each Lie superalgebra in the above list admits an odd simple root system, and the odd principal $f$ is given by the sum of the simple root vectors. For the explicit odd simple root systems and their Dynkin diagrams, see \cite[Section 6]{Song24free}. Moreover, the superconformality of the SUSY W-algebra implies that of the principal W-algebras.

\begin{cor}
    Let $\g$ be one of the Lie superalgebras in the list \eqref{eq: classification}. Then, the corresponding principal W-algebra of level $k\neq-h^{\vee}$ has a superconformal vector of central charge
    \begin{equation} \label{eq: central charge}
        \frac{k\, \textup{sdim}(\g)}{k+h^{\vee}}-3k(H|H)-\sum_{\alpha\in I_+}(-1)^{p(\alpha)}(12m_{\alpha}^2-12m_{\alpha}+2)-\frac{1}{2}\textup{sdim}(\g_{\frac{1}{2}}),
    \end{equation}
    where $[H,u_{\alpha}]=2 m_{\alpha} u_{\alpha}$ for each $\alpha\in I_+$.
\end{cor}
\begin{proof}
   From Corollary \ref{cor: isom for all} and \cite[Theorem 4.6]{Song24free}, one directly obtains that the principal W-algebra is superconformal with central charge
   \begin{equation} \label{eq: susy central charge}
    \frac{k\, \textup{sdim}(\g)}{k+h^{\vee}}+\frac{1}{2}\textup{sdim}(\g)+12 \sum_{\alpha\in I_+}(-1)^{p(\alpha)}m_{\alpha}-3 \textup{sdim}(\mathfrak{n}_+)-3(k+h^{\vee})(H|H).
\end{equation}
The central charge \eqref{eq: susy central charge} coincides with \eqref{eq: central charge} due to the following two equalities:
\begin{equation} \label{eq: cc proof}
    \textup{sdim}(\g)=2\,\textup{sdim}(\n_+)-\textup{sdim}(\g_{\frac{1}{2}}),\quad
    h^{\vee}(H|H)=4\sum_{\alpha\in I_+}(-1)^{p(\alpha)}m_{\alpha}^2.
\end{equation}
The first equality in \eqref{eq: cc proof} follows from $\textup{sdim}(\g)=\textup{sdim}(\n_+)+\textup{sdim}(\n_-)+\textup{sdim}(\g_0)$ and $\g_0=[f,\g_{\frac{1}{2}}]$. The second equality in \eqref{eq: cc proof} is obtained by taking the bilinear form $(\cdot |H)$ on both sides of $\sum_{\alpha\in \widetilde{I}}(-1)^{p(\alpha)}[v^{\alpha},[v_{\alpha},H]]=2h^{\vee}H$, where $\{v^{\alpha}\}_{\alpha\in \widetilde{I}}$ and $\{v_{\alpha}\}_{\alpha\in \widetilde{I}}$ are dual bases of $\g$ satisfying $(v^{\alpha}|v_{\beta})=\delta_{\alpha,\beta}$.
\end{proof}

\begin{rem}
    The central charge in \eqref{eq: central charge} coincides with that of the Virasoro field in \cite[Theorem 2.2]{KRW03}. When $f$ is non-principal, the first equality in \eqref{eq: cc proof} should be replaced by $\textup{sdim}(\g)=2\textup{sdim}(\mathfrak{n}_+)+\textup{sdim}(\g_0^f)- \textup{sdim}(\g_{\frac{1}{2}})$, since in general $\g_0=\g_0^f \oplus [f,\g_{\frac{1}{2}}]$. Thus, for arbitrary $f$, the central charges \eqref{eq: susy central charge} and \eqref{eq: central charge} differ by $\frac{1}{2}\textup{sdim}(\g_0^f)$.
\end{rem}

Note that an explicit formula for the superconformal vector in an ordinary W-algebra remains unknown, since its expression in the SUSY W-algebra in terms of building blocks has not yet been described. We conclude this section with the examples of $\mathfrak{osp}(1|2)$ and $\mathfrak{sl}(2|1)$, the two smallest Lie superalgebras in the list \eqref{eq: classification}.

\begin{example} \label{ex: affine osp}
Consider $\g=\mathfrak{osp}(1|2)=\textup{Span}_{\C}\{E,e,H,f,F\}$. Note that $h^{\vee}=3/2$ and $\g^f=\C F$. As a vertex algebra, the principal SUSY W-algebra $W^k(\bar{\g},f)$ is freely generated by $\omega(\bar{F})$ and $D\omega(\bar{F})$, where
\begin{equation*}
    \begin{aligned}
    \omega(\bar{F})=&J_{\bar{F}}-\frac{2k+3}{4}DJ_{\bar{f}}-\frac{1}{2}:\!J_{\bar{H}}J_{\bar{f}}\!:+\frac{2k+3}{8}:\!J_{\bar{H}}(DJ_{\bar{H}})\!:+\frac{(k+1)(2k+3)}{4}\partial J_{\bar{H}}
    \end{aligned}
\end{equation*}
for the building blocks $J_{\bar{\g}_{\leq 0}}$. The SUSY Miura map images of the fields are
\begin{equation*}
    \begin{aligned}
    \bar{\mu}^k(\omega(\bar{F}))=& \frac{(2k+3)^{\frac{3}{2}}}{8}:\!\Phi J^{(h)}\!:+\frac{(k+1)(2k+3)^{\frac{3}{2}}}{4}\partial \Phi,\\
    \bar{\mu}^k(D\omega(\bar{F}))=& \frac{2k+3}{8}:\!J^{(h)}J^{(h)}\!:-\frac{(2k+3)^{2}}{8}:\! \Phi(\partial \Phi)\!:+\frac{(k+1)(2k+3)}{4}\partial J^{(h)},
    \end{aligned}
\end{equation*}
where $\Phi:=\frac{1}{\sqrt{2k+3}}J_{\bar{H}}$ forms the neutral free fermion algebra, and $J^{(h)}:=DJ_{\bar{H}}$ forms the Heisenberg vertex algebra.

According to \cite[Section 8.2]{KW04}, the principal nonSUSY W-algebra $W^k(\g,F)$ is freely generated by $J^{\{f_{\alpha}\}}$ and $J^{\{f_{2\alpha}\}}$, whose Miura map images are
\begin{equation*}
    \begin{aligned}
    \mu^k(J^{\{f_{\alpha}\}})=&\frac{1}{2}:\!\Phi J^{(h)}\!:+(k+1)\partial \Phi,\\
    \mu^k(J^{\{f_{2\alpha}\}})=&-\frac{1}{4}:\!J^{(h)}J^{(h)}\!:-\frac{k+1}{2}\partial J^{(h)}+\frac{2k+3}{4}:\!\Phi(\partial \Phi)\!:
    \end{aligned}
\end{equation*}
Therefore, by mapping
\begin{equation*}
    \omega(\bar{F})\mapsto \frac{(2k+3)^{\frac{3}{2}}}{4} J^{\{f_{\alpha}\}}, \quad D\omega(\bar{F})\mapsto -\frac{2k+3}{2} J^{\{f_{2\alpha}\}},
\end{equation*}
we get an isomorphism $W^k(\bar{\g},f)\simeq W^k(\g,F)$.
\end{example}

\begin{example} Consider $\g=\mathfrak{sl}(2|1)=\textup{Span}_{\C}\{E,H,F,e,f,\tilde{e},\tilde{f},U\}$, where $\g_{\bar{0}}=\textup{Span}_{\C}\{E,H,F,U\}$ and $\g_{\bar{1}}=\textup{Span}_{\C}\{e,f,\tilde{e},\tilde{f}\}$. The elements $(E,H,F,e,f)$ and $(-E,H,-F,\tilde{e},\tilde{f})$ form $\mathfrak{osp}(1|2)$ subalgebra, respectively. Also, we have $[f, \tilde{f}]=[e,\tilde{e}]=0$ and $[e,\tilde{f}]=[\tilde{e},f]=U$. Note that $h^{\vee}=1$ and $\g^f=\C \tilde{f}\oplus \C F$. As a vertex algebra, the principal SUSY W-algebra $W^k(\bar{\g},f)$ is freely generated by $\omega(\bar{\tilde{f}})$, $D\omega(\bar{\tilde{f}})$, $\omega(\bar{F})$, and $D\omega(\bar{F})$, where
    \begin{equation*}
        \begin{aligned}
            \omega(\bar{\tilde{f}})=&J_{\bar{\tilde{f}}}-(k+1)DJ_{\bar{U}}+\frac{1}{2}:\!J_{\bar{U}}J_{\bar{H}}\!:,\\
            \omega(\bar{F})=&J_{\bar{F}}-\frac{k+1}{2}DJ_{\bar{f}}-\frac{1}{2}:\!J_{\bar{f}}J_{\bar{H}}\!:+\frac{1}{2}:\!J_{\bar{\tilde{f}}}J_{\bar{U}}\!:\\            
            &+\frac{k+1}{4}:\!(DJ_{\bar{H}})J_{\bar{H}}\!:-\frac{k+1}{4}:\!(DJ_{\bar{U}})J_{\bar{U}}\!:+\frac{(k+1)^2}{2}\partial J_{\bar{H}}.
        \end{aligned}
    \end{equation*}
    Take the simple coroots $h_1$ and $h_2$ satisfying $(h_1|h_1)=(h_2|h_2)=0$ and $(h_1|h_2)=1$. Then, $h_i$'s can be written as $h_1=\frac{1}{2}(H-U)$ and $h_2=\frac{1}{2}(H+U)$. Denote by $\Phi_1:=\frac{1}{\sqrt{k+1}}J_{\bar{h}_i}$, $\Phi_2:=-\frac{1}{\sqrt{k+h^{\vee}}}J_{h_2}$, and $J^{(h_i)}:=DJ_{\bar{h}_i}$ for $i=1,2$. Then, the SUSY Miura map images are
    \begin{equation*}
        \begin{aligned}
            \bar{\mu}^k(\omega(\bar{\tilde{f}}))=&(k+1)(J^{(h_1)}-J^{(h_2)})+(k+1):\!\Phi_1 \Phi_2\!:, \\
            \bar{\mu}^k(D\omega(\bar{\tilde{f}}))=&(k+1)^{\frac{3}{2}}(\partial \Phi_1+\partial \Phi_2)+\sqrt{k+1}:\!\Phi_1 J^{(h_2)}\!:+\sqrt{k+1}:\!\Phi_2 J^{(h_1)}\!:,\\
            \bar{\mu}^k(\omega(\bar{F}))=&\frac{(k+1)^{\frac{3}{2}}}{2}\big(:\!\Phi_1 J^{(h_2)}\!:-:\!\Phi_2 J^{(h_1)}\!:\big)+\frac{(k+1)^{\frac{5}{2}}}{2}(\partial \Phi_1-\partial \Phi_2),\\
            \bar{\mu}^k(D\omega(\bar{F}))=&(k+1):J^{(h_1)}J^{(h_2)}:+\frac{(k+1)^2}{2}(:\!\Phi_1(\partial \Phi_2)\!:+:\!\Phi_2(\partial \Phi_1)\!:)+\frac{(k+1)^2}{2}(\partial J^{(h_1)}+\partial J^{(h_2)})
        \end{aligned}
    \end{equation*}
    Recall that the principal nonSUSY W-algebra $W^k(\g,F)$ is freely generated by $J,G^+,G^-$, and $L$, whose explicit formulas are given in \cite[Section 7]{KRW03}. As in Example \ref{ex: affine osp}, one can find the isomorphism $W^k(\bar{\g},f)\rightarrow W^k(\g,F)$ as
    \begin{equation*}
        \begin{gathered}
        \omega(\bar{\tilde{f}})\mapsto (k+1)J, \quad D\omega(\bar{F})\mapsto (k+1)^2 L,\\
        D\omega(\bar{\tilde{f}})\mapsto (k+1)^{\frac{3}{2}}G^+ -\sqrt{k+1}G^-,\quad  \omega(\bar{F})\mapsto -\frac{(k+1)^{\frac{5}{2}}}{2}G^+-\frac{(k+1)^{\frac{3}{2}}}{2}G^-
        \end{gathered}
    \end{equation*}
    by comparing their Miura map images.
\end{example}

\section{Zhu algebras of SUSY W-algebras} \label{sec:Zhu}

\subsection{Complexes with good almost linear differentials}
This section reviews some useful facts on cohomologies with good, almost linear differentials. See \cite{DK06} for the details.

Let $\g$ be a finite-dimensional vector superspace 
with the grading 
\begin{equation} \label{eq:grading on g}
    \g= \bigoplus_{\substack{p,q\in \frac{\mathbb{Z}}{2},\, p+q\in \mathbb{Z}_+;\\ \Delta\in \frac{\mathbb{Z}_{>0}}{2}}}\g^{pq}[\Delta]
\end{equation}
and $\mathcal{T}(\g)$ be the tensor algebra with the induced grading $\mathcal{T}(\g)=\bigoplus_{p,q\in \frac{\mathbb{Z}}{2}, \Delta\in \frac{\mathbb{Z}_{>0}}{2}} \mathcal{T}^{pq}[\Delta].$
Suppose $\g$ is a nonlinear Lie superalgebra, i.e., there is a bilinear skew-symmetric bracket with the Jacobi identity satisfying  
\begin{equation} \label{eq: non-linearity of g}
    [\g^{p_1 q_1}[\Delta_1],\g^{p_2 q_2}[\Delta_2]]\subset \bigoplus_{l\geq 0, \Delta<(\Delta_1+\Delta_2)} \mathcal{T}(\g)^{p_1+p_2+l, q_1+q_2-l}[\Delta].
\end{equation}
For such $\g$, we have the $\mathbb{Z}_+$-grading called charge and two filtrations defined  by 
\begin{equation} \label{eq:Lie_filtration, grading}
     \g= \bigoplus_{n\in \mathbb{Z}_+} \g^n, \ \quad  F^p(\g)= \bigoplus_{p'\geq p} \g^{pq}, \ \quad \g_\Delta= \bigoplus_{\Delta'\leq \Delta} \g[\Delta'],
\end{equation}
where $\g^n= \bigoplus_{p+q=n, \, \Delta \in \mathbb{Z}_{>0}/2} \g^{pq}$ and $\g[\Delta]= \bigoplus_{p,q\in \mathbb{Z}/2} \g^{pq}[\Delta]$. Then \eqref{eq:Lie_filtration, grading} naturally induces the corresponding $\mathbb{Z}_+$-grading and filtrations on the universal enveloping algebra $U(\g).$ 

Suppose we have an almost linear good differential $d:U(\g)\to U(\g)$, increasing the charge by $1$ and preserving both filtrations. Here, the almost linear good differential means that the graded differential $d^{\text{gr}}$ has the properties 
\begin{equation}
\begin{aligned}
       &  d^{\text{gr}}(\g^{p q})\subset \g^{p \, q+1} \text{ for any } p,q \in \mathbb{Z}/2;\\
       & \text{Ker} (d^{\text{gr}}|_{\g^{pq}}) = \text{Im}(d^{\text{gr}}|_{\g^{p\,q-1}}) \text{ for } p+q \neq 0.
\end{aligned}
\end{equation}
For the definition of the graded differential $d^{\text{gr}}$, we refer to \cite[Section 4]{DK06}.

\begin{thm} [\cite{DK06}] \label{thm:cohomology-Lie algebra}
    Let $\g$ be a non-linear Lie superalgebra with the properties \eqref{eq:grading on g}, \eqref{eq: non-linearity of g}, \eqref{eq:Lie_filtration, grading} and $d$ is a good almost linear differential on $U(\g).$ Then the followings hold.
    \begin{enumerate}[(1)]
        \item The cohomology $H(U(\g),d)$ is concentrated on the charge $0$ part, that is, 
        \begin{equation*}
            H(U(\g),d)=H^0(U(\g),d).
        \end{equation*} 
        \item Let $H(\g,d)$ be the non-linear Lie superalgebra spanned by elements $\{E_i=\epsilon_i+ \eta_i \mid i\in I\}\subset U(\g)$, where $E_i$ is in $F^{p_i}(U(\g))$, $\{\epsilon_i|i\in I\}\subset \g$ is a basis of $H(\g,d^{\text{gr}})$ and $E_i-e_i\in F^{p_i+1/2}(U(\g)).$ Then 
        \begin{equation*}
         H(U(\g),d)\simeq U(H(\g,d)).
        \end{equation*}
    \end{enumerate}
\end{thm}

Consider the $\mathbb{C}[\partial]$-module $R=\mathbb{C}[\partial]\otimes \g$ with the grading 
\begin{equation}\label{eq:grading on R}
     R= \bigoplus_{\substack{p,q\in \frac{\mathbb{Z}}{2}, \, p+q\in \mathbb{Z}_+;\\ \Delta\in \frac{\mathbb{Z}_{>0}}{2}}}R^{pq}[\Delta]
\end{equation}
induced from that of $\g$. It follows from the fact that $\partial$ preserves the $(p,q)$ bi-grading and increases the $\Delta$-grading by $1$. Suppose that $R$ is a nonlinear Lie conformal algebra, i.e., there is a bilinear skew-symmetric bracket satisfying the Jacobi identity and 
\begin{equation} \label{eq: non-linearity of R}
    R^{p_1 q_1}[\Delta_1]_{(n)}R^{p_2 q_2}[\Delta_2]\subset \bigoplus_{l\geq 0} \mathcal{T}(R)^{p_1+p_2+l, q_1+q_2-l}[\Delta_1+\Delta_2-n-1].
\end{equation}
We define $\mathbb{Z}_+$-grading called charge, decreasing filtration, and  $\mathbb{Z}_{>0}/2$-grading as
\begin{equation} \label{eq:LCA_filtration, grading}
     R= \bigoplus_{n\in \mathbb{Z}_+} R^n, \ \quad  F^p(R)= \bigoplus_{p'\geq p} R^{pq}, \ \quad R[\Delta]= \bigoplus_{p,q} R^{pq}[\Delta],
\end{equation}
where $R^n= \bigoplus_{p+q=n, \, \Delta \in \mathbb{Z}_{>0}/2} R^{pq}$.  Analogously, we define the corresponding gradings and filtration on the universal enveloping vertex algebra $V(R).$ Then we get the vertex algebra analog of Theorem \ref{thm:cohomology-Lie algebra}.

\begin{thm}[\cite{DK06}] \label{thm:cohomology-LCA}
     Let $R=\mathbb{C}[\partial]\otimes \g$ be a non-linear Lie conformal algebra with the properties \eqref{eq:grading on R}, \eqref{eq: non-linearity of R}, \eqref{eq:LCA_filtration, grading} and $d$ is a good almost linear differential on $V(R).$ Then the followings hold.
    \begin{enumerate}[(1)]
        \item The cohomology $H(V(R),d)$ is concentrated on the charge $0$ part, that is,
        \begin{equation*}
            H(V(R),d)=H^0(V(R),d).
        \end{equation*} 
        \item Let $H(R,d)$ be the non-linear Lie conformal algebra $\mathbb{C}[\partial]$-spanned by elements $\{E_i=\epsilon_i+ \eta_i \mid i\in I\}\subset V(R)$ where $E_i$ is in $F^{p_i}(V(R))$, $\{\epsilon_i|i\in I\}\subset R$ is a $\mathbb{C}[\partial]$-basis of $H(R,d^{\text{gr}})$ and $E_i-e_i\in F^{p_i+1/2}(V(R)).$ Then 
        \begin{equation*}
         H(V(R),d)\simeq V(H(R,d)).
        \end{equation*}
    \end{enumerate}
\end{thm}

Let $H$ be a Hamiltonian operator on $V(R)$ such that $H(a)=\Delta(a) a$ for all $a\in V(R)[\Delta(a)].$  The $H$-twisted Zhu algebra $Zhu_H(V(R))$ is the associative algebra 
\[Zhu_H(V(R)):= V(R)/J(R),\]
where $J(R)$ is the two-sided ideal of $V(R)$ generated by $(\partial+H)(V(R)).$ Denote the image of $a\in V(R)$ in $Zhu_H(V(R))$ by $Zhu_H(a).$
The Lie bracket and the associative product on $Zhu_H(V(R))$ is defined by 
\begin{equation} \label{eq:Zhu_products}
\begin{aligned}
    & [Zhu_H(a),Zhu_H(b)]:=[Zhu_H(a),Zhu_H(b)]_\hbar \big|_{\hbar=1},\\
    & Zhu_H(a)Zhu_H(b)= Zhu_H(:\!ab\!:) + \int^{1}_0  [Zhu_H(H(a)),Zhu_H(b)]_x dx,
\end{aligned}
\end{equation} where
$=[Zhu_H(a),Zhu_H(b)]_\hbar=\sum_{j\in \mathbb{Z}_+} { \Delta_a - 1 \choose j} \, \hbar^j\, Zhu_H(a_{(j)} b).$

\begin{thm}[\cite{DK06}]\label{thm:Zhu-coho}
    Let $R$ and $d$ be given as in Theorem \ref{thm:cohomology-LCA}.
    Then
    \begin{equation*}
        Zhu_H (H(V(R),d)) \simeq H(Zhu_H(V(R)), Q)
    \end{equation*}
    as associative algebras, where $Q$ is the differential on $Zhu_H(V(R))$ induced from $d.$ Moreover, $Zhu_H(V(R))\simeq U(\g)$ so that 
    \begin{equation*}
         Zhu_H (H(V(R),d))\simeq  H(U(\g),Q).
    \end{equation*}
\end{thm}

\subsection{Zhu algebras of SUSY W-algebras} \label{subsec:SUSY finite}
In the following, we adopt the assumptions and notations for a Lie superalgebra $\g$ as stated in Section \ref{sec: FF of SUSY W-alg} and let $k\neq- h^{\vee}$.

Recall the SUSY BRST complex 
\begin{equation} 
  C^k(\bar{\g},f):=V^{k}(\bar{\g})\otimes F(\bar{\mathfrak{A}})
\end{equation}
and its superconformal vector $G=\omega+\tau+\partial \bar{H}$, where 
$\omega$ is the Kac-Todorov vector and 
\begin{equation} \label{eq:FF_conformal}
\begin{aligned}
        \tau = & \sum_{\alpha\in I_+} (-1)^{p(\alpha)}2 j_{\alpha}:(\partial\phi_{\alpha})\phi^{\alpha}: \\
        & -\sum_{\alpha\in I_+} (-1)^{p(\alpha)}(1-2 j_{\alpha}):\phi_{\alpha}\partial \phi^{\alpha}:+\sum_{\alpha\in I_+}:(D\phi_{\alpha})(D\phi^{\alpha}): 
\end{aligned}
\end{equation}
for $j_\alpha$ such that $[x,u_\alpha]= j_\alpha u_\alpha.$ Then the conformal weights given by $G$ on $C^k(\bar{\g},f)$ are 
\begin{equation}
    \Delta(\bar{u}_\alpha)=\Delta(\phi_\alpha)=\frac{1}{2}-j_\alpha, \quad \Delta(\phi^\alpha)= j_\alpha, \quad \Delta(D)=\frac{1}{2}.
\end{equation}
Denote the induced Hamiltonian operator $H$ on $C^k(\bar{\g},f)$ and write $H(a)=\Delta(a) a$ for any homogeneous element $a\in C^k(\bar{\g},f).$

\begin{defn} \label{def:Zhu of W}
    We denote the cohomology of the complex $Zhu_H 
    (C^k(\bar{\mathfrak{g}},f))$ with the differential $Q$ induced from $d_{(0|0)}$ by 
    \[ W^{\textit{fin}}(\bar{\mathfrak{g}},f):= H(Zhu_H(C^k(\bar{\mathfrak{g}},f)), Q).\]
\end{defn}

The Zhu algebra $Zhu_H(C^k(\bar{\mathfrak{g}},f))$ is the associative superalgebra 
\[Zhu_H(C^k(\bar{\mathfrak{g}},f)) \simeq U(\widetilde{\g}_*\oplus\phi^{\widetilde{\n}_-}\oplus  \phi_{\widetilde{\n}}),\]
where $\widetilde{\g}_*= \g \oplus \bar{\g}$, $\widetilde{\n}= \n \oplus \bar{\n}$ and $\widetilde{\n}_-= \n_- \oplus \bar{\n}_-$ and the commutator relation of the nonlinear Lie superalgebra $\widetilde{\g}_*\oplus\phi^{\widetilde{\n}_-}\oplus  \phi^{\widetilde{\n}}$ is given by 
\begin{equation} \label{eq: Lie bracket zhu brst}
    \begin{aligned}
         & [\bar{a},\bar{b}]_* = (k+h^\vee)(a|b), \quad [\bar{a},b]_*=(-1)^{p(a)p(b)}\overline{[a,b]}, \quad  [a,b]_*=(-1)^{p(a)p(b)}[a,b] , \\
         & [\phi^{m}, \phi_n]_*=(-1)^{p(m)}[\phi^{\bar{m}},\phi_{\bar{n}}]_*=(m|n), \quad [\phi^{\bar{m}},\phi_n]_*=[\phi^{m},\phi_{\bar{n}}]_*=0,
    \end{aligned}
\end{equation}
for $a,b\in \g$, $n\in \n$ and $m\in \n_-.$ In \eqref{eq: Lie bracket zhu brst}, we denote $\boldsymbol{\bar{a}}:=Zhu_H(\bar{a}),$ 
\begin{equation} \label{eq:finite a}
     \boldsymbol{a}:=Zhu_H(D\bar{a})-(k+h^{\vee})(x|a),
\end{equation}
 $\boldsymbol{\phi}^{\bar{m}}:= Zhu_H(\phi^{\bar{m}}),$ $\boldsymbol{\phi}^m:= Zhu_H(D\phi^{\bar{m}})$, $\boldsymbol{\phi}_{n}:=Zhu_H(\phi_{n}),$ and $\boldsymbol{\phi}_{\bar{n}}:=Zhu_H(D\phi_{n})$ with an abuse of notation. To avoid confusion, we denote the bracket on $\widetilde{\g}_*$ by $[\ , \ ]_*$ and the bracket on $\g$ by $[\ , \ ]$.

The differential $Q$ on $Zhu_H C^k(\bar{\g},f)$ is given to satisfy the following equalities:
\begin{equation} \label{eq:Q on C}
    \begin{aligned}
        & Q(\boldsymbol{\bar{a}})=\sum_{\beta\in I_+} \Big[\, (-1)^{p(\bar{a})p(\beta)}\boldsymbol{\phi}^{\bar{u}^\beta}\overline{\boldsymbol{[u_\beta,a]}}+(-1)^{p(\beta)+1}(k+h^\vee)\boldsymbol{\phi}^{u^\beta}(u_\beta|a)\, \Big],\\
        & Q(\boldsymbol{a})=\sum_{\beta\in I_+} \Big[\,(-1)^{p(\bar{a})p(\beta)+1}\boldsymbol{\phi}^{u^\beta}\overline{\boldsymbol{[u_\beta,a]}}+(-1)^{p(a)p(\beta)}\boldsymbol{\phi}^{\bar{u}^\beta}\boldsymbol{[u_\beta,a]}\, \Big],\\
        & Q(\boldsymbol{\phi}^{\bar{u}^\alpha})=\frac{1}{2}\sum_{\alpha,\beta\in I_+}\Big[\,(-1)^{p(\bar{\alpha})p(\beta)}\boldsymbol{\phi}^{\bar{u}^\beta} \boldsymbol{\phi}^{\overline{[u_\beta,u^\alpha]}}\, \Big],\\
        & Q(\boldsymbol{\phi}^{u^\alpha})=\frac{1}{2}\sum_{\alpha,\beta\in I_+}\Big[\,(-1)^{p(\bar{\alpha})p(\beta)+1}\boldsymbol{\phi}^{u^\beta}\boldsymbol{\phi}^{\overline{[u_\beta,u^\alpha]}}+(-1)^{p(\alpha)p(\beta)}\boldsymbol{\phi}^{\bar{u}^\beta}\boldsymbol{\phi}^{[u_\beta,u^\alpha]}\, \Big],\\
        & Q(\boldsymbol{\phi}_{u_\alpha}) = (-1)^{p(\bar{\alpha})}\boldsymbol{\bar{u}_\alpha}-(f|u_\alpha)+\sum_{\beta\in I_+}(-1)^{p(\bar{\alpha})p(\beta)}\boldsymbol{\phi}^{\bar{u}^\beta}\boldsymbol{\phi}_{[u_\beta,u_\alpha]},\\
        & Q(\boldsymbol{\phi}_{\bar{u}_\alpha})= (-1)^{p(\alpha)}\boldsymbol{u_\alpha} \\
        & \hskip 1.5cm + \sum_{\beta\in I_+}\Big[\,(-1)^{p(\bar{\alpha})p(\beta)+1}\boldsymbol{\phi}^{u^\beta}\boldsymbol{\phi}_{[u_\beta,u_\alpha]}+ (-1)^{p(\alpha)p(\beta)}\boldsymbol{\phi}^{\bar{u}^\beta}\boldsymbol{\phi}_{\overline{[u_\beta,u_\alpha]}}\, \Big]\ ,
    \end{aligned}
\end{equation}
where $\Delta(\phi^{\beta})=j_\beta.$ Remark that if we denote $\widehat{\boldsymbol{a}}:=Zhu_H(D\bar{a})$ (cf. $a$ in \eqref{eq:finite a}), then 
\begin{equation}
    [\widehat{\boldsymbol{a}},\widehat{\boldsymbol{b}}]_*=(-1)^{p(a)p(b)}\big(\widehat{\boldsymbol{[a,b]}} - (k+h^\vee)([x,a]|b)\big)
\end{equation}
and
\begin{equation}
Q(\widehat{\boldsymbol{a}})=\sum_{\beta\in I_+} \Big[\,(-1)^{p(\bar{a})p(\beta)+1}\boldsymbol{\phi}^{u^\beta}\boldsymbol{\overline{[u_\beta,a]}}+(-1)^{p(a)p(\beta)}\boldsymbol{\phi}^{\bar{u}^\beta}\widehat{\boldsymbol{[u_\beta,a]}}-(-1)^{p(\beta)}(k+h^\vee) j_\beta \boldsymbol{\phi}^{\bar{u}^\beta}(u_\beta|a)\, \Big].
\end{equation}

\vskip 2mm

Now, recall from \eqref{eq: building block SUSY} the building blocks $J_{\bar{a}}$'s in $C^k(\bar{\g},f)$. Again, by abusing notation, let us denote $\bold{J}_{\bar{a}}:= Zhu_H(J_{\bar{a}})$ and $\bold{J}_{a}:= Zhu_H(DJ_{\bar{a}})$. Then one can check that the subspaces
$r_+:= \bold{J}_{\widetilde{\n}}\oplus \boldsymbol{\phi}_{\widetilde{\n}}$ and $r_-:= \bold{J}_{\widetilde{\g}_{\leq 0}} \oplus \boldsymbol{\phi}^{\widetilde{n}_-}$ of $Zhu_H C^k(\bar{\g},f)$ are both nonlinear Lie superalgebras. For example, the commutation relations in $r_-$ are given by 
\begin{equation}
    \begin{aligned}
        & [ \bold{J}_{\bar{a}},  \bold{J}_{\bar{b}}]_*=(k+h^\vee)(a|b), \quad [ \bold{J}_{\bar{a}},  \bold{J}_{b}]_*=(-1)^{p(a)p(b)} \bold{J}_{\overline{[a,b]}}, \\
        & [ \bold{J}_{a},  \bold{J}_b]_*= (-1)^{p(a)p(b)}\big(  \bold{J}_{[a,b]}-(k+h^\vee)([x,a]|b) \big),\\
        & [ \bold{J}_{\bar{a}},  \boldsymbol{\phi}^{n}]_*= (-1)^{p(a)}[ \bold{J}_{a},  \boldsymbol{\phi}^{\bar{n}}]_*= -\boldsymbol{\phi}^{\overline{[n,a]}},\quad [ \bold{J}_{a},  \boldsymbol{\phi}^{n}]_*= -\boldsymbol{\phi}^{[n,a]}
    \end{aligned}
\end{equation}
and 
$[ \bold{J}_{\bar{a}},  \boldsymbol{\phi}^{\bar{n}}]= [   \boldsymbol{\phi}^{\bar{n}_1},  \boldsymbol{\phi}^{\bar{n}_2}]=  [  \boldsymbol{\phi}^{n_1},  \boldsymbol{\phi}^{\bar{n}_2}]= [  \boldsymbol{\phi}^{n_1},  \boldsymbol{\phi}^{{n}_2}]
=
0$ for $a\in \g_{\leq 0}$ and $n, n_1, n_2\in \n_-$. 

\begin{prop} \label{prop:finite_r+-}
    We have the following properties for the differential $Q$ on the complex $Zhu_H C^k(\bar{\g},f)$:
    \begin{enumerate}[(1)]
        \item $Q(U(r_+))\subset U(r_+)$ and $Q(U(r_-))\subset U(r_-)$,
        \item $U(\bar{\g},f)\simeq H(U(r_-),Q).$
    \end{enumerate}
\end{prop}
\begin{proof}
(1) 
Recall from \cite{MRS21} the action of $d_{(0|0)}$ on $J_{\bar{a}}\in C^k(\bar{\g},f)$ for $a\in \g_{\leq 0}$ as follows:
\begin{equation}
\begin{aligned}
    d_{(0|0)}(J_{\bar{a}}) & =\sum_{\beta\in I_+}(-1)^{p(\bar{a})p(\beta)} :\!\phi^\beta \, \big(\, J_{\pi_{\leq}\overline{[u_\beta,a]}}+(f|[u_\beta,a]) \,  \big)\!: \\
    & + \sum_{\beta\in I_+}(-1)^{p(\beta)+1}(k +h^\vee) D\phi^\beta(u_\beta|a).
\end{aligned}
\end{equation}
Since $D$ and $d_{(0|0)}$ supercommute, we have 
\begin{equation*}
\begin{aligned}
    d_{(0|0)}(DJ_{\bar{a}}) = & \sum_{\beta\in I_+} (-1)^{p(\bar{a})p(\beta)+1} :\!D(\phi^\beta) \, \big(\, J_{\pi_{\leq}\overline{[u_\beta,a]}}+(f|[u_\beta,a]) \,  \big)\!:  \\ & +  \sum_{\beta\in I_+}(-1)^{p(a)p(\beta)}:\!\phi^\beta DJ_{\pi_{\leq}\overline{[u_\beta,a]}}\!: + \sum_{\beta\in I_+}(-1)^{p(\beta)}(k+h^{\vee}) \partial\phi^\beta(u_\beta|a).
\end{aligned}
\end{equation*}
Now, by \eqref{eq:Zhu_products}, the differential $Q$ acts on the elements in $r_-$ as

\begin{align}
     &
     \begin{aligned}
    Q(  \bold{J}_{\bar{a}})  =&  \sum_{\beta\in I_+}(-1)^{p(\bar{a})p(\beta)}  \boldsymbol{\phi}^{\bar{u}^\beta}\, \big( \,   \bold{J}_{\pi_{\leq}\overline{[u_\beta,a]}} + (f|[u_\beta,a]) \, \big)\\ & + \sum_{\beta\in I_+}(-1)^{p(\beta)+1}(k +h^\vee)  \boldsymbol{\phi}^{u^\beta}(u_\beta|a), \label{eq:Q & Ja} 
     \end{aligned}
     \\
     &\begin{aligned}
     Q(  \bold{J}_a)=& \sum_{\beta\in I_+}(-1)^{p(a)p(\beta)} \big( \,   \boldsymbol{\phi}^{\bar{u}^\beta}\boldsymbol{J}_{\pi_{\leq}[u_\beta,a]} -(k+h^{\vee})   \boldsymbol{\phi}^{\bar{u}^\beta}([x,u_\beta]|a)\, \big) \label{eq:Q & DJa-1}  \\ 
    &  + \sum_{\beta\in I_+}(-1)^{p(\bar{a})p(\beta)+1}  \boldsymbol{\phi}^{u^\beta} \, \big( \, \bold{J}_{\pi_{\leq}\overline{[u_\beta,a]}} +(f|[u_\beta,a])\, \big).
     \end{aligned}
\end{align}
The terms in \eqref{eq:Q & DJa-1} follow from the definition of the product on the Zhu algebra. 
More precisely, we know $H(\phi^\beta)= j_\beta \phi^\beta$, $H(D\phi^\beta)= (j_\beta+\frac{1}{2}) D\phi^\beta$ and 
\begin{equation}
        [  \boldsymbol{\phi}^{u^\beta},   \bold{J}_{\pi_{\leq }\overline{[u_\beta,a]}}]_*=(-1)^{p(\beta)+1}  \boldsymbol{\phi}^{[\pi_{\leq}[u_\beta,a],u^{\beta}]}, \quad  [  \boldsymbol{\phi}^{\bar{u}^\beta},   \bold{J}_{\pi_{\leq }[u_\beta,a]}]_*=- \boldsymbol{\phi}^{[\pi_{\leq}[u_\beta,a],u^{\beta}]}.
\end{equation}
Hence,
\begin{equation}
\begin{aligned}
& Zhu_H\Big((-1)^{p(\bar{a})p(\beta)+1}:D(\phi^\beta) J_{\pi_{\leq}\overline{[u_\beta,a]}}:+(-1)^{p(a)p(\beta)}:\phi^\beta DJ_{\pi_{\leq}\overline{[u_\beta,a]}}:\Big) \\
& = (-1)^{p(\bar{a})p(\beta)+1}  \boldsymbol{\phi}^{u^\beta}   \bold{J}_{\pi_{\leq}\overline{[u_\beta,a]}}+(-1)^{p(a)p(\beta)}  \boldsymbol{\phi}^{\bar{u}^\beta}   \bold{J}_{\pi_{\leq}[u_\beta,a]}.
\end{aligned}
\end{equation}
By \eqref{eq:Q on C}, \eqref{eq:Q & Ja} and \eqref{eq:Q & DJa-1}, we conclude $Q(U(r_-))\subset U(r_-).$ Also, since 
\begin{equation} \label{eq:Q r+}
    Q(\boldsymbol{\phi}_{u_\alpha})= (-1)^{p(\bar{\alpha})}\bold{J}_{\bar{u}_\alpha}-(f|u_\alpha), \quad Q(\boldsymbol{\phi}_{\bar{u}_\alpha})= (-1)^{p(\alpha)}\bold{J}_{u_\alpha},
\end{equation}
we have $Q(U(r_-))\subset (U(r_-)).$\\
(2) By \eqref{eq:Q r+}, 
one can easily see that $H(U(r_+),Q)=\mathbb{C}.$ Observe $Zhu_H C^k(\bar{\g},f) = U(r_+)\otimes U(r_-)$, and hence 
\begin{equation*}
  W^{\textit{fin}}(\bar{\g},f)\simeq H(U(r_+),Q)\otimes H(U(r_-),Q)\simeq H(U(r_-),Q).
\end{equation*}
\end{proof}

Recall that the SUSY W-algebra $W^k(\bar{\g},f)$ can also be defined via the subcomplex $\widetilde{C}^k(\bar{\g},f)$ generated by $r_-$. Since we have $Zhu_H \widetilde{C}^k(\bar{\g},f) = U(r_-),$ Proposition \ref{prop:finite_r+-} implies that
\begin{equation*}
    W^{\textit{fin}}(\bar{\g},f)\simeq H(Zhu_H \widetilde{C}^k(\bar{\g},f) ,Q).
\end{equation*}
The following theorem shows that the SUSY finite W-algebra is also considered as the Zhu algebra of the corresponding SUSY affine W-algebra.

\begin{thm} \label{def:SUSY finite W}
As associative superalgebras,  we have 

\begin{equation}
 W^{\textit{fin}}(\bar{\g},f)\simeq Zhu_H \, W^k(\bar{\g},f)\simeq H^0(Zhu_H \widetilde{C}^k(\bar{\mathfrak{g}},f),Q).
\end{equation}   
\end{thm}

\begin{proof}
Recall that the conformal weight on $\widetilde{C}^k(\bar{\g},f)$ is defined by 
\begin{equation}
    \Delta(J_{\bar{u}_\alpha})=\frac{1}{2}-j_\alpha, \quad \Delta(\phi^\alpha)= j_\alpha, \quad \Delta(D)=\frac{1}{2}
\end{equation}
for $u_\alpha \in \g(j_\alpha).$
   Consider the $\frac{1}{2}\Z$ bi-grading on $\widetilde{C}^k(\bar{\g},f)$ defined as follows:
\begin{equation} \label{eq:bigrading}
 \text{gr}(J_{\bar{u}_\alpha})=(j_\alpha, -j_\alpha), \quad \text{gr}(\phi^\alpha)=\Big(-j_\alpha+\frac{1}{2}, j_\alpha+\frac{1}{2}\Big), \quad \text{gr}(D)=(0,0).
\end{equation}
Then,
\begin{equation*}
d_{(0|0)}^{\text{gr}}(J_{\bar{a}})= \sum_{\beta\in I_+}(-1)^{p(\bar{a})p(\beta)} \phi^\beta(f|[u_\beta,a]),\quad  d_{(0|0)}^{\text{gr}}(\phi^{\alpha})=0    
\end{equation*}
for any $\alpha\in I_+$, and hence 
\begin{equation*}
H(\widetilde{C}^k(\bar{\g},f), d_{(0|0)}^{\text{gr}})= V(\mathbb{C}[\partial]\otimes (J_{\bar{\g}^f}\oplus DJ_{\bar{\g}^f})).
\end{equation*}
This implies that the differential $d_{(0|0)}$ is a good almost linear differential. Hence, by Theorem \ref{thm:Zhu-coho}, we have 
\[ H(Zhu_H \widetilde{C}^k(\bar{\g},f),Q)\simeq Zhu_H H(\widetilde{C}^k(\bar{\g},f),d_{(0|0)}) = Zhu_H W^k(\bar{\g},f). \]

Now, we consider the $\frac{1}{2}\Z$-grading $\Delta$ on $U(r_-)$ such that 
    \begin{equation}
    \Delta(J_{\bar{u}_\alpha})=\frac{1}{2}-j_\alpha, \quad \Delta(J_{u_\alpha})=1-j_\alpha,\quad  \Delta(\phi^{\bar{u}^\alpha})= j_\alpha, \quad \Delta(\phi^{u^\alpha})=\frac{1}{2}+j_\alpha.
\end{equation}
and the $\frac{1}{2}\Z$ bi-grading given by 
\begin{equation} \label{eq:bigrading_finite}
    \text{gr}(J_{\bar{u}_\alpha})= \text{gr}(J_{u_\alpha})=(j_\alpha, -j_\alpha), \quad \text{gr}(\phi^{\bar{u}^\alpha})=\text{gr}(\phi^{u^\alpha})=\Big(-j_\alpha+\frac{1}{2}, j_\alpha+\frac{1}{2}\Big).
\end{equation}
Then, using the same argument as in the affine case, the differential $Q$ is a good almost linear differential. Hence, $Zhu_H W^k(\bar{\g},f)\simeq H^0(Zhu_H \widetilde{C}^k(\bar{\g},f),Q).$
\end{proof}

\section{finite SUSY W-algebras} \label{sec:finite W}
In this section, we introduce finite SUSY W-algebras and compare them with the Zhu algebras of SUSY W-algebras. Throughout the rest of this paper, we set $\g$ to be a finite-dimensional simple basic Lie superalgebra satisfying the assumptions stated in Section \ref{sec: FF of SUSY W-alg}.

\subsection{Definition of finite SUSY W-algebras}
Recall that $\bar{\g}$ is the parity reversed vector superspace of $\g.$ Consider the Lie superalgebra $\widetilde{\g}=\g\oplus \bar{\g}$ 
endowed with the Lie bracket extending that of $\g$ and 
\begin{equation} \label{eq: Lie bracket Takiff}
    [\bar{a},b]= \overline{[a,b]}, \quad [\bar{a},\bar{b}]=0
\end{equation}
for $a,b\in \g$. Note that $\widetilde{\g}$ is isomorphic to the SUSY Takiff algebra $\g\otimes \wedge(\theta)$ under the map $a \mapsto a$ and $\bar{a}\to a\otimes \theta$ for $a\in \g.$

Using the bilinear form  $( \ | \ )$ on $\g$, define the bilinear form $\left<\ |\  \right>$ on $\widetilde{\g}$ by
\begin{equation}
    \left<\bar{a}|b \right>=(-1)^{p(a)}\left< a|\bar{b}\right>= (a|b), \quad \left<a|b\right> = \left< \bar{a}|\bar{b}\right>=0.
\end{equation}
This is a supersymmetric nondegenerate invariant bilinear form on $\widetilde{\g}.$

One can extend the Lie bracket \eqref{eq: Lie bracket Takiff} by considering the central extension $\widetilde{\g}\oplus \mathbb{C}K$ 
of $\widetilde{\g}$. The Lie bracket on $\widetilde{\g}\oplus \mathbb{C}K$ extends that of $\g$ and 
\begin{equation}
    [\bar{a},b]= \overline{[a,b]}, \quad [\bar{a},\bar{b}]=(-1)^{p(a)}K(a|b)
\end{equation}
for $a,b\in \g$. Let us denote
\begin{equation}                U^k(\widetilde{\g}):=U(\widetilde{\g}\oplus\mathbb{C}K)/U(\widetilde{\g}\oplus\mathbb{C}K)( K-(k+h^\vee))
\end{equation}
for $k\neq -h^\vee$ in $\mathbb{C}.$

\begin{rem} \label{rem:isom of affines}
   The SUSY affine vertex algebra $V^k(\bar{\g})$ with the $\Lambda$-bracket \eqref{eq:SUSY_affine_nosign} has 
    the Hamiltonian operator $H$ coming  from  Kac-Todorov superconformal vector in  $ V^k(\bar{\g})$ (\cite{KT85,HK07}) and it gives rise to the Zhu algebra
   \begin{equation}
       Zhu_{H}\, V^k(\bar{\g})\simeq U^k(\widetilde{\g}).
   \end{equation}
    In addition, by the Remark \ref{rem:isom of affines_sect3}, we can also conclude $U^k(\widetilde{\g})$ is the Zhu algebra of the SUSY affine vertex algebra given by \eqref{eq: susy affine bracket}.
\end{rem}

Denote $\widetilde{\n}:=\n \oplus \bar{\n}$ and let $\chi$ be the character on $\widetilde{\n}$ defined by 
\begin{equation} \label{eq:chi}
     \chi(\widetilde{n})= \left<f|\widetilde{n}\right>.
\end{equation}
 Then $\chi(n)=0$ and $\chi(\bar{n})= -(f|n)$ for $n\in \n$. Consider the $1$-dimensional  $\widetilde{\n}$-module
$\mathbb{C}_{\pm\chi}$   given by $\widetilde{n}\cdot 1 = \pm \left<f|\widetilde{n}\right>$. Then, $U^k(\widetilde{\g})$ is a right $\widetilde{\n}$-module via the right multiplication and $U^k(\widetilde{\g})\otimes_{\widetilde{\n}}\mathbb{C}_\chi$ is a left $\widetilde{\n}$-module by the adjoint action.  

\begin{defn}  \label{def:finite W-alg}
The finite SUSY W-algebra is an associative superalgebra 
\begin{equation}
    U(\widetilde{\g},f)= ( \, U^k (\widetilde{\g})\otimes_{\widetilde{\n}}\mathbb{C}_{-\chi}\, )^{\textup{ad}\, \widetilde{\n}}.
\end{equation}
Equivalently, it is the associative superalgebra given by
\begin{equation}
    U(\widetilde{\g},f)=\Big( \, U^k (\widetilde{\g})\, /\, U^k (\widetilde{\g}) (\widetilde{n}+\left<f|\widetilde{n}\right>|\, \widetilde{n}\in \widetilde{\n}\, )\Big)^{\textup{ad}\, \widetilde{\n}}.
\end{equation}
Note that this is the same superalgebra introduced as a finite SUSY W-algebra in \cite{CCS24+}.
\end{defn}

\begin{rem} \label{eq:rem_ell_indep}
     For a nonzero constant $\ell$ and the corresponding character $\chi_{\ell}$ of $\widetilde{\n}$ satisfying $\chi_{\ell}(\widetilde{n})= \ell\left< f|\widetilde{n}\right>,$ let us consider $ U_{\ell}(\widetilde{\g},f): = ( \, U^k (\widetilde{\g})\otimes_{\widetilde{\n}}\mathbb{C}_{-\chi_{\ell}}\, )^{\textup{ad}\, \widetilde{\n}}.$
Then 
\begin{equation} \label{eq: character indepen}
    U_{\ell}(\widetilde{\g},f) \simeq   U(\widetilde{\g},f).
\end{equation}   
Indeed, the algebra automorphism of $U^k(\widetilde{\g})$ given by 
$\widetilde{a} \mapsto \ell^{- 2 j_a} \widetilde{a}$
for $\widetilde{a}\in \g_{j_a} \oplus \bar{\g}_{j_a}$ induces the isomorphism between  $U(\widetilde{\g},f)$ and  $U_{\ell}(\widetilde{\g},f)$ in \eqref{eq: character indepen}. We also note that the finite SUSY W-algebra $U(\widetilde{\g},f)$ is independent of the choice of $k$. This can be deduced from the automorphism on $U^k(\widetilde{\g})$ given by $\bar{a}\mapsto \frac{1}{\sqrt{k+h^{\vee}}}\bar{a}, \ a\mapsto a$ for $a\in \g$.

\end{rem}

Obviously, $U(\widetilde{\g},f)$ is the degree $0$ part of the Lie superalgebra cohomology $ H(\widetilde{\n}, U^k (\widetilde{\g})\otimes_{\widetilde{\n}}\mathbb{C}_{-\chi}).$ In the rest of this section, we show that the cohomology is concentrated in degree $0$ part, which further implies that  $U(\widetilde{\g},f)\simeq H(\widetilde{\n}, U^k (\widetilde{\g})\otimes_{\widetilde{\n}}\mathbb{C}_{-\chi})$. As an associative superalgebra, we can identify $U^k(\widetilde{\g})\otimes_{\widetilde{\n}}\mathbb{C}_
{-\chi}$ with the universal enveloping algebra $U^k(\widetilde{\mathfrak{p}})$ of the nonlinear Lie subalgebra $\widetilde{\mathfrak{p}}=\mathfrak{p}\oplus\bar{\mathfrak{p}}$ of $\widetilde{\g}$, where $\g=\n\oplus \mathfrak{p}$. We also consider the exterior algebra $\wedge\, \widetilde{\n}^*$ generated by $\widetilde{\n}^*$. Note that this space can be viewed as the universal enveloping algebra $U(\Pi\,  \widetilde{\n}^*)$ of the supercommutative Lie superalgebra $\Pi\, \widetilde{\n}^*$, where $\Pi\, \widetilde{\n}^*$ is the parity reversed space of $\widetilde{\n}^*$. The complex of the Lie superalgebra cohomology of $U^k(\widetilde{\g})\otimes \mathbb{C}_
{-\chi}$  is
\begin{equation} \label{eq:complex_Lie}
       C(\widetilde{\n}, U^k (\widetilde{\g})\otimes_{(\widetilde{\n})}\mathbb{C}_
{-\chi}):=\wedge\, \widetilde{\n}^* \otimes U^k(\widetilde{\mathfrak{p}}), 
\end{equation}
that is, the universal enveloping algebra of the nonlinear lie superalgebra $\Pi\, \widetilde{\n}^*\oplus \widetilde{\mathfrak{p}}$. The differential $d_L$ on $C(\widetilde{\n}, U^k (\widetilde{\g})\otimes_{\widetilde{\n}}\mathbb{C}_{-\chi})$ is defined as
\begin{equation} \label{eq:differential_Lie}
\begin{aligned}
    & d_L(1\otimes a)= \sum_{\alpha\in I_+} u_\alpha^* \otimes \text{ad}u_\alpha (a)+ \sum_{\alpha\in I_+} \bar{u}_\alpha^* \otimes \text{ad}\bar{u}_\alpha (a), \\
    & d_L(\widetilde{u}_\gamma^*\otimes 1)\\
    &= \frac{1}{2} \sum_{\alpha,\beta\in I_+} \big( c_{\alpha,\widetilde{\gamma}}^{\beta} u_\alpha^* \wedge u_\beta^* + c_{\alpha,\widetilde{\gamma}}^{\bar{\beta}}u_\alpha^* \wedge \bar{u}_\beta^* + c_{\bar{\alpha},\widetilde{\gamma}}^{\beta} \bar{u}_\alpha^*\wedge u_\beta^* + c_{\bar{\alpha},\widetilde{\gamma}}^{\bar{\beta}}\bar{u}_\alpha^*\wedge\bar{u}_\beta^*\big),
\end{aligned}
\end{equation}
where $\pi_{+}([u_\alpha,\widetilde{u}_\gamma])= \sum_{\beta\in I_+}(c_{\alpha,\widetilde{\gamma}}^{\beta}u_\beta + c_{\alpha,\widetilde{\gamma}}^{\bar{\beta}}\bar{u}_\beta)$ and  $\pi_{+}([\bar{u}_\alpha,\widetilde{\gamma}])= \sum_{\beta\in I_+}(c_{\bar{\alpha},\widetilde{\gamma}}^{\beta}u_\beta + c_{\bar{\alpha},\widetilde{\gamma}}^{\bar{\beta}}\bar{u}_\beta)$ for the canonical projection $\pi_+: \widetilde{\g}\oplus \mathbb{C}K \to \widetilde{\n}$ and $\widetilde{u}_\gamma= u_\gamma$ or $\bar{u}_\gamma.$ Furthermore, for an arbitrary element $a_1 \wedge \cdots \wedge a_s \otimes m$ in the complex, the differential $d_L$ acts as follows:
 \begin{equation} \label{eq:differencil_arb}
 \begin{aligned}
     & d_{L}(a_1 \wedge \cdots \wedge a_s \otimes m) \\
     & = \sum_{i=1}^s(-1)^{p(a_1)+\cdots +p(a_{i-1})+i-1}a_1\wedge \cdots \wedge d_L(a_i)\wedge \cdots \wedge a_s \otimes m\\
     & \qquad \qquad + (-1)^{p(a_1)+\cdots +p(a_{s})+s}a_1 \wedge \cdots \wedge a_s \otimes d_L(m).
\end{aligned}
 \end{equation}

On the complex \eqref{eq:complex_Lie}, we consider the $\frac{1}{2}\Z$-grading $\Delta$ given by 
    \begin{equation}
    \Delta(1\otimes \bar{u}_\alpha)=\frac{1}{2}-j_\alpha, \quad \Delta(1\otimes u_\alpha)=1-j_\alpha,\quad  \Delta(\bar{u}_\alpha^*\otimes 1)= \frac{1}{2}+j_\alpha, \quad \Delta(u_\alpha^*\otimes 1)=j_\alpha
\end{equation}
and the $\frac{1}{2}\Z$ bi-grading called $(p,q)$-grading provided as
\begin{equation} \label{eq:bigrading_finite_Lie}
\begin{aligned}
       & \text{gr}(1\otimes \bar{u}_\alpha)= \text{gr}(1\otimes u_\alpha)=(j_\alpha, -j_\alpha), \\ & \text{gr}(\bar{u}_\alpha^*\otimes 1)=\text{gr}(u_\alpha^*\otimes 1)=\Big(-j_\alpha+\frac{1}{2}, j_\alpha+\frac{1}{2}\Big).
\end{aligned}
\end{equation}
With respect to the $(p,q)$-grading, the graded differential $d_L^{\text{gr}}$ satisfies 
\begin{equation}
    d_L^{\text{gr}}(1\otimes a)=\sum_{\alpha\in I_+}\Big(\left<f|[u_\beta,a] \right>u_\alpha^* + \left<f|[\bar{u}_\alpha,a] \right>\bar{u}_\alpha^*\Big) \otimes 1, \quad  d_L^{\text{gr}}(\widetilde{u}_\beta^*\otimes 1)=0
\end{equation}
for $a\in \widetilde{\mathfrak{p}}$ and $\widetilde{u}_\beta=u_\beta$ or $\bar{u}_\beta$. Hence, $d_L^{\text{gr}}$ is a good almost linear differential. By Theorem \ref{thm:cohomology-Lie algebra}, $H\big(C(\widetilde{\n}, U^k(\widetilde{\g})\otimes_{\widetilde{\n}} \mathbb{C}_
{-\chi}), d_L\big)=H^0\big(C(\widetilde{\n}, U^k(\widetilde{\g})\otimes_{\widetilde{\n}} \mathbb{C}_
{-\chi}), d_L\big)$ which directly implies the following proposition.

\begin{prop} The finite SUSY W-algebra $U(\widetilde{\g},f)$ is isomorphic to the Lie superalgebra cohomology $H(\widetilde{\n}, U^k(\widetilde{\g})\otimes_{\widetilde{\n}} \mathbb{C}_
{-\chi}).$
    
\end{prop}

\subsection{$U(\widetilde{\g},f)$ and $Zhu_H W^k(\bar{\mathfrak{g}},f)$ }

In order to compare $U(\widetilde{\g},f)$ and $Zhu_H W^k(\bar{\mathfrak{g}},f)$, let us derive another construction of $U(\widetilde{\g},f)$ via a double complex. Consider the left $\widetilde{\n}$-modules $U(\widetilde{\n})_{-\chi}$ and $U(\widetilde{\n})$, both of which are isomorphic to  $U(\widetilde{\n})$ as vector superspaces. For $\widetilde{n}\in \widetilde{\n}$ and $A\in U(\widetilde{\n})$, the actions of $\widetilde{n}$ on $A$ are defined as $\widetilde{n}\cdot A=(\widetilde{n}-\left<f|\widetilde{n}\right>)A$ in the former and by the left multiplication in the latter. Similarly, let $U^k(\widetilde{\g})$ and $U^k(\widetilde{\g})_\chi$ be the right $\widetilde{\n}$-modules via the right multiplication and the right action on $U^k(\widetilde{\g})_\chi\simeq  U^k(\widetilde{\g})\otimes \mathbb{C}_\chi,$ respectively. Define
\begin{equation}
    M:= U^k(\widetilde{\g})\otimes_{U(\widetilde\n)}U(\widetilde{\n})_{-\chi}\simeq U^k(\widetilde{\g})_{\chi}\otimes_{U(\widetilde\n)}U(\widetilde{\n})
\end{equation}
to be the $\widetilde{\n}$-bimodule whose left action is the adjoint action and the right action is the sign twisted multiplication. Here, the sign twisted means that $m\cdot \widetilde{n} =(-1)^{p(\widetilde{n})} m\widetilde{n} $ for $m\in M$ and $\widetilde{n}\in \widetilde{\n}.$ 

The Lie superalgebra homology of the right $\widetilde{\n}$-module $M$ is defined via the complex 
\begin{equation}
    C_h:=M\otimes \wedge\, \widetilde{\n}.
\end{equation} 
 The differential $d_h$ on $C_h$ satisfies
 \begin{equation}
 \begin{aligned}
         d_h(m\otimes {c_1}\wedge {c_2}\cdots \wedge {c_s}) & =\sum_{1\leq t \leq s} (-1)^{P_1}m\cdot c_t \otimes {c_1}\wedge \cdots \widecheck{c_t}\cdots \wedge {c_s} \\
         & + \sum_{1\leq r<t \leq s}(-1)^{P_2}m  \otimes {[c_r,c_t]}\wedge {c_1}\wedge \cdots  \widecheck{c_r}\cdots  \widecheck{c_t} \cdots \wedge {c_s},
 \end{aligned}
 \end{equation}
where $P_1=(p(c_t)+1)\big(p(c_1)+\cdots+p(c_{t-1})+t-1\big)$ and $P_2=(p(c_r)+p(c_t))\big(p(c_1)+\cdots+p(c_{r-1})+r-1\big)+(p(c_t)+1)(p(c_{r+1})+\cdots+p(c_{t-1})+(t-r-1))$. In particular, we have 
 \begin{equation}
     d_h( 1\otimes \widetilde{n})= (-1)^{\widetilde{n}}(\widetilde{n}+\left<f|\widetilde{n} \right>).
 \end{equation}
 Now, we consider the Lie superalgebra cohomology of the left $\widetilde{\n}$-module $M\otimes  \wedge\, \widetilde{\n}$. The complex for the cohomology is
 \begin{equation}
     C_c:= \wedge\, \widetilde{\n}^* \otimes (M\otimes \wedge\,\widetilde{\n})
 \end{equation}
 and the differential $d_c$ is defined as in  \eqref{eq:differential_Lie} and \eqref{eq:differencil_arb}.
Combining the $\widetilde{\n}$-module homology and cohomology, consider the complex 
 \begin{equation}
     C_{I}=\wedge\, \widetilde{\n}^* \otimes M\otimes \wedge\,\widetilde{\n}
 \end{equation}
 and the differential $d_{I}= d_c+(-1)^{\delta}\otimes d_h$, where $\delta$ reflects the grading of $ \wedge\, \widetilde{\n}^*$ part. Note that one can check that $d_{I}^2=0$ by direct computation.

 \begin{prop}
     The following two associative superalgebras are isomorphic:
     \[U(\widetilde{\g},f)\simeq H(C_{I},d_{I}).\]
 \end{prop}
\begin{proof}
    The proposition follows directly from the analogous statement in the proof of Theorem A.6 in \cite{DK06}. Here, we briefly review the idea in the reference. Let us consider the bicomplex 
    \[ C_I= \bigoplus_{p,-q\in \Z_+} C^{p,q}, \quad d_I= d_c+d_h, \]
    where $C^{p,q}=\wedge^{p} \, \widetilde{\n}^* \otimes M \otimes \wedge^{-q}\, \widetilde{\n}$. In order to find the cohomology $H(C_I,d_I),$ we consider the corresponding spectral sequence $(E_r,\, d^r\, | \, r\geq 1).$ Then,
    \begin{equation}
    \begin{aligned}
         E_1^{p,q} & = H^{p,q}(C_I, d_c) = \wedge^p\, \widetilde{\n}^* \otimes H^q((U^k(\widetilde{\g})_\chi\otimes_{\widetilde{\n}} U(\widetilde{\n}))\otimes \wedge\, \widetilde{\n}, d_h) \\
         & = \wedge^p\, \widetilde{\n}^* \otimes U^k(\widetilde{\g})_\chi\otimes_{\widetilde{\n}} H^q(U(\widetilde{\n})\otimes \wedge\, \widetilde{\n}, d_h) \\
         & = \delta_{q,0} \wedge^p\, \widetilde{\n}^* \otimes (U^k(\widetilde{\g})_\chi\otimes_{U(\widetilde{\n})} \mathbb{C}) \\
         & = \delta_{q,0} C^p(\widetilde{\n}, U^k(\widetilde{\g})\otimes_{U(\widetilde{\n})} \mathbb{C}_{-\chi}).
    \end{aligned}
    \end{equation}
    Hence, the spectral sequence stabilizes at $r=2$ and $E_{\infty}\simeq H(C_I, d_I) \simeq H(\widetilde{n}, U^k(\widetilde{\g})\otimes_{U(\widetilde{\n})} \mathbb{C}_{-\chi}) = U^k(\widetilde{\g},f)$.
\end{proof}

Recall that we have a nondegenerate bilinear form $\left< \ | \ \right>$ on $\widetilde{\g}.$ We can identify $\widetilde{\n}_-$ with $\widetilde{\n}^*$ using the bilinear form. More precisely, since $\left<\bar{u}^\alpha|u_\beta \right>=(-1)^{p(\alpha)}\left<u^\alpha|\bar{u}_\beta \right> =\delta_{\alpha,\beta}$, we can identify
\begin{equation} \label{eq:ident}
    u_\beta^* \equiv \bar{u}^\beta,  \quad \bar{u}_\beta^*\equiv (-1)^{p(\beta)} u^{\beta}.
\end{equation}  
Hence, the coadjoint action of $\widetilde{\n}$ on $\widetilde{\n}^*$ gives rise to the action of  $\widetilde{\n}$ on $\widetilde{\n}_-$. The coadjoint actions are given as follows:
\begin{equation} \label{eq:coad}
\begin{aligned}
        &  \bar{u}_\alpha \cdot u_\beta^*(\widetilde{n})= (-1)^{p(\bar{\alpha})p(\beta)+1} u_\beta^*([\bar{u}_\alpha,\widetilde{n}])=(-1)^{p(\bar{\alpha})p(\beta)+1} \left<\bar{u}^\beta\,|\,[\bar{u}_\alpha,\widetilde{n}]\right>=0, \\
        & u_\alpha \cdot u_\beta^*(\widetilde{n})=(-1)^{p(\alpha)p(\beta)+1}u_\beta^*([u_\alpha,\widetilde{n}])=\left< \overline{[u_\alpha,u^\beta]}\, |\, \widetilde{n} \right>,\\
        & u_\alpha \cdot \bar{u}_\beta^*(\widetilde{n})=(-1)^{p(\alpha)p(\bar{\beta})+1}\bar{u}_\beta^*([u_\alpha,\widetilde{n}])= (-1)^{p(\alpha)+p(\beta)}\left< [u_\alpha, u^\beta]\,|\,\widetilde{n}\right>,\\
        & \bar{u}_\alpha \cdot \bar{u}_\beta^*(\widetilde{n})=(-1)^{p(\bar{\alpha})p(\bar{\beta})+1}\bar{u}_\beta^*([\bar{u}_\alpha,\widetilde{n}])= (-1)^{p(\alpha)+p(\beta)+1}\left< \overline{[u_\alpha, u^\beta]}\,|\, \widetilde{n} \right>,
\end{aligned}
\end{equation}
for any $\widetilde{n}\in \widetilde{\n}.$ Therefore, $\bar{u}_\alpha \cdot \bar{u}^\beta= 0$ and
\begin{equation}
\begin{aligned}
     &   u_\alpha \cdot \bar{u}^\beta= \pi_{<0}\big(\, \overline{[u_\alpha, u^\beta]}\, \big),\\
     &  u_\alpha \cdot u^\beta = (-1)^{p(\alpha)}\pi_{<0}\big([u_\alpha,u^\beta]\big), \\
     &  \bar{u}_\alpha \cdot u^\beta=(-1)^{p(\alpha)+1}\pi_{<0}\big(\, [u_\alpha, u^\beta]\, \big),
\end{aligned}
\end{equation}
where $\pi_{<0}: \widetilde{\g}\to \widetilde{\n}_-$ is the canonical projection map.

Consider the sets $\Psi^{\widetilde{\n}_-}=\{\Psi_{ \widetilde{n}_-}\,|\,\widetilde{n}_-\in \widetilde{\n}_-\}$ and $\Psi_{\widetilde{\n}}=\{\Psi_{n}\,|\,n\in \widetilde{\n}\}$. Here, the parity of the elements are given by $p(\Psi^{\widetilde{n}_-})=p(\widetilde{n}_-)$ and $p(\Psi_{\widetilde{n}})=p(\widetilde{n})$. Take the complex  
\begin{equation}
    C_{II}= S(\Psi^{\widetilde{\n}_-})\otimes M \otimes S(\Psi_{\widetilde{\n}})
\end{equation}
and the bijection map $\iota: C_I \to C_{II}$ defined by
\begin{equation} \label{eq:iota}
    a_1 \wedge \cdots \wedge a_t \otimes m \otimes c_1 \wedge \cdots \wedge c_t  \mapsto  \Psi^{a_1} \cdots \Psi^{a_t} \otimes m \otimes \Psi{c_1} \cdots \Psi_{c_t}.
\end{equation}
Here, we used the identification \eqref{eq:ident} in the map $a_i \in \widetilde{\n}^* \ \mapsto\ \Psi^{a_i}  \in \Psi^{\widetilde{\n}_-},$ that is,
\begin{equation}
    u_\alpha^* \mapsto \Psi^{\bar{u}^\alpha}=:\Psi^{\bar{\alpha}}, \quad  \bar{u}_\alpha^* \mapsto (-1)^{p(\alpha)}\Psi^{u^\alpha}=:(-1)^{p(\alpha)}\Psi^{\alpha}.
\end{equation}
Obviously, the endomorphism 
\begin{equation}
    d_{II}:= \iota\circ d_{I}\circ\iota^{-1}
\end{equation}
is a differential on $C_{II}$. In particular, we have
\begin{equation} \label{eq:diff_psi}
\begin{aligned}
        & d_{II}(\Psi^{\bar{\beta}} \otimes (1\otimes1)\otimes 1)= \frac{1}{2}\sum_{\alpha\in I_+}\Psi^{\bar{\alpha}}\Psi^{\overline{[u_\alpha,u^\beta]}}\otimes (1\otimes1)\otimes 1,\\
    & d_{II}(\Psi^{\beta} \otimes (1\otimes1)\otimes 1)= \frac{1}{2}\sum_{\alpha\in I_+}\Big((-1)^{p(\alpha)}\Psi^{\bar{\alpha}}\Psi^{[u_\alpha, u^\beta]}-\Psi^{\alpha}\Psi^{\overline{[u_\alpha,u^\beta]}}\Big) \otimes (1\otimes1)\otimes 1,\\
    & d_{II}( 1\otimes (m\otimes1) \otimes 1)= \sum_{\alpha\in I_+}\Big( \Psi^{\bar{\alpha}}\otimes [u_\alpha,m] +(-1)^{p(\alpha)}\Psi^{\alpha}\otimes \overline{[u_\alpha,m]}\Big)\otimes 1,\\
     & d_{II}( 1\otimes (\overline{m}\otimes 1) \otimes 1)= \sum_{\alpha\in I_+} \Big((-1)^{p(\alpha)}\Psi^{\bar{\alpha}}\otimes \overline{[u_\alpha,m]} + \Psi^{\alpha}\otimes (k+h^{\vee})(u_\alpha|a)\Big)\otimes 1,\\
      & d_{II}(1\otimes (1\otimes1)\otimes \Psi_{\bar{n}} )= (-1)^{p(n)} 1\otimes (1\otimes n)\otimes 1 + \sum_{\alpha\in I_+}\Psi^{\bar{\alpha}}\otimes (1\otimes1)\otimes \Psi_{\overline{[u_\alpha,n]}} \\
      &  \hskip 5cm +\sum_{\alpha\in I_+}(-1)^{p(\alpha)}\Psi^{\alpha}\otimes (1\otimes1)\otimes  \Psi_{[u_\alpha,n]},\\
    & d_{II}(  1\otimes (1\otimes1)\otimes \Psi_{n})=(-1)^{p(\bar{n})} 1\otimes (1\otimes \bar{n})\otimes 1\\
      &  \hskip 5cm +\sum_{\alpha\in I_+}(-1)^{p(\alpha)}\Psi^{\bar{\alpha}}\otimes (1\otimes1)\otimes \Psi_{[u_\alpha,n]}
\end{aligned}
\end{equation}
for $\beta\in I_+$, $m\in \g$, $n\in \n$. Also, we define  $\Psi^{a}:=\Psi^{\pi_{<0}(a)}$, $\Psi^{\bar{a}}:=\Psi^{\overline{\pi_{<0}(a)}}$, $\Psi_{a}:=\Psi_{\pi_{>0}(a)}$ and $\Psi_{\bar{a}}:=\Psi_{\overline{\pi_{>0}(a)}}$ for any $a\in \g$. By construction, the following proposition is obtained directly.

\begin{prop}
    The two complexes $(C_I,d_I)$ and $(C_{II}, d_{II})$ are quasi-isomorphic by the map $\iota$ in \eqref{eq:iota}. In other words, 
    \[ U(\widetilde{\g},f)\simeq H(C_{II},d_{II}).\]
\end{prop}

Recall that $Zhu_H(W^k(\widetilde{\g},f))\simeq H(Zhu_H(C^k(\bar{\g},f), Q)),$ where $Zhu_H(C^k(\bar{\g},f))=U(\widetilde{\g}_*\oplus \phi^{\widetilde{\n}_-} \oplus \phi_{\widetilde{\n}})$ and $Q$ is given by \eqref{eq:Q on C}.

\begin{thm} \label{thm:finite and Zhu}
   The complex $(Zhu_H\, C^k(\bar{\g},f),Q)$ is quasi-isomorphic to $(C_{II}, d_{II}).$ Hence, as associative superalgebras,
    \[ U(\widetilde{\g},f)\simeq Zhu_H W^k(\bar{\g},f).\]
\end{thm}
\begin{proof}
    By Remark \ref{eq:rem_ell_indep}, it is enough to show that $U_{\ell}(\widetilde{\g},f)\simeq Zhu_H W^k(\bar{\g},f)$ for any nonzero constant $\ell.$
  Let $\ell=-\sqrt{-1}$.    Let us consider the following elements in $C_{II}$:
\begin{equation}  \label{eq:new psi}
    \begin{aligned}
    \Psi^{\mathsf{n}}_*:=\sqrt{-1}^{p(\mathsf{n})}\Psi^{\mathsf{n}},\quad a_*&:= \sqrt{-1}^{p(a)}a, \quad \Psi_{n*}:=\sqrt{-1}^{p(n)} \Psi_{n}, \\
    \Psi^{\bar{\mathsf{n}}}_*:=\sqrt{-1}^{p(\mathsf{n})}\Psi^{\bar{\mathsf{n}}},\quad \bar{a}_*&:= \sqrt{-1}^{p(a)}\bar{a}, \quad \Psi_{\bar{n}*}:=\sqrt{-1}^{p(n)} \Psi_{\bar{n}},
    \end{aligned}
\end{equation}
where $\mathsf{n}\in \n_-,$ $a\in \g,$ and $n\in \n$. In Remark \ref{rem:isom of affines_sect3}, one can see where the coefficients in \eqref{eq:new psi} originate from. Define a map $\iota: C_{II}\to Zhu_H C^k(\bar{\g},f)$ by 
    \begin{equation}        
    \begin{aligned}
& \Psi^{\bar{\alpha}}_*\otimes (1\otimes 1) \otimes 1\mapsto(-1)^{p(\alpha)} \boldsymbol{\phi}^{\bar{u}^\alpha},\quad  \Psi^{\alpha}_*\otimes (1\otimes 1) \otimes 1\mapsto(-1)^{p(\bar{\alpha})}  \boldsymbol{\phi}^{u^\alpha}  \\
 & 1\otimes \big(\tilde{a}_*\otimes 1\big)\otimes 1 \mapsto  \boldsymbol{\tilde{a}}\\
 & 1\otimes (1\otimes 1)\otimes \Psi_{n*} \mapsto \boldsymbol{\phi}_{n},\quad 1\otimes (1\otimes 1)\otimes \Psi_{\bar{n}*} \mapsto \boldsymbol{\phi}_{\bar{n}}, \\
        \end{aligned}
    \end{equation}
for $\tilde{a}\in \widetilde{\mathfrak{p}},$ $\tilde{n}\in \widetilde{\n}$ and $n\in\n.$ It is obvious that $\iota$ is bijective and we have to show that $\iota\circ d_{II}= Q\circ \iota$. Since the differentials $d_{II}$ and $Q$ are both odd derivations, it is enough to prove 
\begin{equation} \label{eq:aim_quasi-isom}
    \iota\circ d_{II}(A)= Q\circ \iota(A)
\end{equation}
for any $A\in \Psi^{\widetilde{\mathsf{n}}}\oplus (\g\otimes 1) \oplus \Psi_{\widetilde{\n}}.$
This equality can be proved directly using   \eqref{eq:Q on C} and \eqref{eq:diff_psi}. For example,
\begin{equation*}
\begin{aligned}
       & \iota \circ d_{II}(\Psi_{n*})  = \iota \circ d_{II}(\sqrt{-1}^{p(n)}\Psi_{n})\\
       & = \sqrt{-1}^{p(n)}\iota \Big(1\otimes (1\otimes (-1)^{p(\bar{n})}\bar{n} )\otimes 1+\sum_{\alpha\in I_+}(-1)^{p(\alpha)}\Psi^{\bar{\alpha}}\otimes (1\otimes1)\otimes \Psi_{[u_\alpha,n]} \Big)\\
         & = \sqrt{-1}^{p(n)} \iota \Big(1\otimes \big(\left((-1)^{p(\bar{n})}\bar{n} + \ell \left< f|\bar{n}\right>\right)\otimes 1\big)\otimes 1 +\sum_{\alpha\in I_+}(-1)^{p(\alpha)}\Psi^{\bar{\alpha}}\otimes (1\otimes1)\otimes \Psi_{[u_\alpha,n]}\ \Big)\\
         & = (-1)^{p(\bar{n})}\boldsymbol{\bar{n}}+ \left<f|\bar{n}\right>+ \sum_{\alpha\in I_+}(-1)^{p(\bar{n})p(\alpha)}\boldsymbol{\phi}^{\bar{u}^\alpha}\boldsymbol{\phi}_{[u_\alpha,n]}= Q(\boldsymbol{\phi}_n)= Q\circ \iota(\Psi_{n *})
    \end{aligned}
\end{equation*}
and 
\begin{equation*}
    \begin{aligned}
        & \iota \circ d_{II}(\Psi^{\beta}_*)  = \iota \circ d_{II}(\sqrt{-1}^{p(\beta)}\Psi^{\beta}) \\
        & = \frac{\sqrt{-1}^{p(\beta)}}{2}  \ \iota \Big(\sum_{\alpha\in I_+}\Big((-1)^{p(\alpha)}\Psi^{\bar{\alpha}}\Psi^{[u_\alpha, u^\beta]}-\Psi^{\alpha}\Psi^{\overline{[u_\alpha,u^\beta]}}\Big) \otimes (1\otimes1)\otimes 1 \Big)\\
        & =  \frac{1}{2}\sum_{\alpha,\beta\in I_+}\Big[\,(-1)^{p(\beta)p(\alpha)+p(\beta)+p(\alpha)}\boldsymbol{\phi}^{u^\alpha}\boldsymbol{\phi}^{\overline{[u_\alpha,u^\beta]}}+(-1)^{p(\beta)p(\alpha)+p(\beta)+1}\boldsymbol{\phi}^{\bar{u}^\alpha}\boldsymbol{\phi}^{[u_\alpha,u^\beta]}\, \Big]\\
        & = Q((-1)^{p(\bar{\beta})}\boldsymbol{\phi}^{u^\beta})=Q\circ \iota(\Psi^{\beta}_*).
    \end{aligned}
\end{equation*}
Here, we used the facts
\begin{equation*}
    \sqrt{-1}^{p(a)}\sqrt{-1}^{p(b)}\sqrt{-1}^{p(ab)}=(-1)^{p(a)p(b)+p(a)+p(b)}\ , \   \big(\sqrt{-1}^{p(a)}\big)^{-1}=(-\sqrt{-1})^{p(a)}.
\end{equation*}
Analogously, one can prove \eqref{eq:aim_quasi-isom} for any other element $A$ in
$\Psi^{\widetilde{\mathsf{n}}}\oplus (\g\otimes 1) \oplus \Psi_{\widetilde{\n}}$. 
\end{proof}

\subsection{Principal SUSY finite W-algebras}

In this section, let $f$ be the odd principal nilpotent element in $\g$ and $F=-\frac{1}{2}[f,f]$. In Theorem \ref{thm:generic_isom} and Corollary \ref{cor: isom for all}, we showed that $W^k(\g,F)\simeq W^k(\g,f)$ for $k\neq -h^{\vee}.$ Now, we want to compare the finite W-algebra $U(\g,F)$ and the SUSY finite W-algebra $U(\widetilde{\g},f)$. Recall the following two facts 
\begin{itemize}
    \item  $U(\g,F)$ is the $H_{N=0}$-twisted Zhu algebra of $W^k(\g,F)$  (\cite{DK06,Genra24}), 
    \item $U(\widetilde{\g},f)$ is the $H_{N=1}$-twisted Zhu algebra of $W^k(\bar{\g},f)$ (Theorem \ref{thm:finite and Zhu}),
\end{itemize}
where $H_{N=0}$ and  $H_{N=1}$ are the Hamiltonian operators induced from the conformal vectors of $W^k(\g,F)$ and $W^k(\widetilde{\g},f),$ respectively. Hence, by comparing two operators $H_{N=0}$ and  $H_{N=1}$, we get the following theorem.

\begin{thm} \label{thm:isom_ finite}
Let $\g$ be a basic simple Lie superalgebra and $f$ be the odd principal nilpotent. For $F=-\frac{1}{2}[f,f]$, we have
    \[ U(\g,F)\simeq U(\widetilde{\g},f). \]
\end{thm}

\begin{proof}
    By the observation in the previous paragraph, it is enough to show that $H_{N=0}$ and  $H_{N=1}$ induce the same operator via the isomorphism between $W^k(\g, F)$ and $W^k(\bar{\g},f)$. More precisely, we need to show that the map $\tau$ in \eqref{eq:domain_screening} preserves the conformal weight. Recall the notations in \eqref{eq:domain_screening}. We know 
    \begin{equation}
        \Delta_0(\Phi_\alpha\otimes 1)=\Delta_1(\bar{h}_\alpha)=\frac{1}{2}, \quad \Delta_0(1\otimes h)= \Delta_1(D\bar{h})=1,
    \end{equation}
    where $\Delta_0$ and $\Delta_1$ are the conformal weights given by $H_{N=0}$ and $H_{N=1}$ and hence we get the theorem.
\end{proof}

\begin{ex} 
Let $\mathfrak{g}=\mathfrak{osp}(1|2)$ spanned by $\{E,H=2x,F,e,f\}\subset \g,$ where $(E,2x,F)$ is the $\mathfrak{sl}_2$-triple and $e,f\in \mathfrak{g}_{\bar{1}}$. Additionally, we fix the Lie brackets 
\begin{equation}
    \begin{aligned}
        & [x,e]=\frac{1}{2}e, \quad [x,f]=-\frac{1}{2}f, \\
        & [e,e]=2E, \quad [f,f]=-2F, \quad [e,f]=[f,e]=-2x, \quad [F,e]=f, \quad [E,f]=e
    \end{aligned}
\end{equation}
and the bilinear form
\begin{equation}
    (E|F)=2(x|x)=1, \quad (e|f)=-(f|e)=-2.
\end{equation}
Let us take the character $\chi$ such that $\bar{e}\mapsto -1, \, e\mapsto 0, \, \bar{E}\mapsto 0, \, E\mapsto 0$ and $k+h^\vee=1.$ Then $U(\widetilde{\g},f)$ is generated by the following two elements
\begin{equation}
    \begin{aligned}
       &  w_{\bar{F}}:= \bar{F}-2\bar{f}\bar{x}+2\bar{x}-f-4\bar{x}x;\\
       & w_{F}:= F-2\bar{f}x-2 f\bar{x}-4x^2,
    \end{aligned}
\end{equation}
and $ \{w_{\bar{F}},w_{F}\}$ is a PBW basis of $U(\widetilde{\g},f).$
Indeed, one can check that $\textup{ad}\, {\bar{e}}\, (w_{\bar{F}})=\textup{ad}\, {e}\, (w_{\bar{F}})=\textup{ad}\, {\bar{e}}\, (w_{F})=\textup{ad}\, {e} \,(w_{F})=0$ in $U^k(\widetilde{\g})/U^k(\widetilde{\g})(\widetilde{n}+\chi(\widetilde{n})\,|\,\widetilde{n}\in \widetilde{\n}).$ For example, $\textup{ad}\, {\bar{e}}\, (w_{F})=0$ can be checked by  
\begin{equation}
    \begin{aligned}
        & \textup{ad}\, {\bar{e}} \,(F)= -\bar{f}\\
        & \textup{ad}\, {\bar{e}}\, (\bar{f}x)= 2x+\bar{f}\big(-\frac{1}{2}\bar{e}\big)=-\frac{1}{2} \bar{f}+2x\\
        & \textup{ad}\, {\bar{e}}\, (f\bar{x}) = -2\bar{x}\bar{x}=-\frac{1}{2}\\
        & \textup{ad}\, {\bar{e}}\,(x^2)= x\big(-\frac{1}{2}\bar{e}\big)+\big(-\frac{1}{2}\bar{e}\big)x=-\frac{1}{2}[\bar{e},x]-x = -x+\frac{1}{4}.
    \end{aligned}
\end{equation}
Similarly, since $\textup{ad}\, {e} \,(F)=-f,$ $\textup{ad}\, {e}\,(\bar{f}x)= 2\bar{x}x,$ $\textup{ad}\, {e}\,(f\bar{x})=-\frac{1}{2}f-2\bar{x}x$, and $\textup{ad}\, {e}\,(x^2)=0$, one can conclude that $w_{F}\in U(\widetilde{\g},f).$ 

Observe that in the commutator $[ w_{\bar{F}}, w_{\bar{F}}]$, the term $[-f,-f]=-2F$ is the only non-constant part contained in the subalgebra generated by $F$ and $\bar{F}$. Hence, $[ w_{\bar{F}}, w_{\bar{F}}]=2w^2_{\bar{F}}=-2w_F + (\text{constant})=:w_F^c.$ Similarly, we can show $[w_{F},w_{\bar{F}}]=[w_{F},w_{F}]=0.$ Therefore, $[w^c_{F},U(\widetilde{\g},f)]=0$ and $U(\widetilde{\g},f)$ can be considered as a supercommutative algebra generated by $w_{\bar{F}}.$

On the other hand, $U(\g,F)$ is known in \cite{Genra24} to be isomorphic to the ghost center $\widetilde{Z}(\g)$ of $\g$, which is the subalgebra generated by
\begin{align*}
    Q = \frac{1}{2}H^2 + EF + FE,\quad
    C = Q + \frac{1}{2}ef - \frac{1}{2}fe
\end{align*}
inside the universal enveloping algebra $U(\g)$ of $\g$. Note that $C$ is the Casimir element of $\g$ and $Q$ is that of $\g_{\bar{0}}$. See the introduction in loc. cit. for the reference of the above facts. Let $T=4Q-4C+\frac{1}{2}$. Then $\widetilde{Z}(\g)=\C[C] \oplus \C[C]T$ with the relation $T^2=4C+\frac{1}{4}$. Under the isomorphism $\widetilde{Z}(\g) \simeq U(\g,F)$, $\C[C]$ is the even part of $U(\g,F)$, while $\C[C]T$ is the odd part. Therefore, we obtain an isomorphism of associative superalgebras:
\begin{align*}
    U(\widetilde{\g},f) \xrightarrow{\sim} \widetilde{Z}(\g) \simeq U(\g,F),\quad
    w_{\bar{F}} \mapsto T,\quad
    w^c_{F} \mapsto 8C+\frac{1}{2}.
\end{align*}

\end{ex}

\section{Conflict of interest statement}
The authors have no conflicts to disclose.
\section{Data availability statement}
The authors declare that the data supporting the findings of this study are available within the paper and the references cited herein.

\bibliographystyle{halpha.bst}
\bibliography{refs}

\end{document}